\documentclass[11pt,epsf]{article}
\usepackage{graphicx,float}
\usepackage{mathrsfs,array,multirow}
\usepackage{amssymb,amsfonts}
\usepackage{amsmath}
\usepackage{color}
\usepackage{mathrsfs}
\usepackage{amsmath}
\usepackage{amssymb}
\usepackage{amstext}
\usepackage{caption,xcolor}
\DeclareCaptionFont{blue}{\color{blue}}
\usepackage[
       colorlinks=true,
      filecolor=black,
       anchorcolor=blue,
      linkcolor=blue,
      citecolor=red,
      urlcolor=blue,
       linktocpage=true,
        plainpages=false,
        breaklinks=true,
            pdfstartview=FitH
           ]{hyperref}

\newcommand{\be}{\begin{equation}}

\newcommand{\ee}{\end{equation}}

\newcommand{\ba}{\begin{array}}

\newcommand{\ea}{\end{array}}

\newcommand{\nb}{\nabla}

\newcommand{\sm}{\sigma}

\newcommand{\ld}{\lambda}

\begin{document}
\begin{titlepage}

\title{\Large \bf Anisotropic Dyonic Black Brane and its Effects on Hydrodynamics}

\vskip 1. cm
\author{
Sunly Khimphun$^{a,}$\footnote{e-mail : kpslourk@sogang.ac.kr},~~
Bum-Hoon Lee$^{a,b,}$\footnote{e-mail : bhl@sogang.ac.kr}, ~~
Chanyong Park$^{b,c,}$\footnote{e-mail : chanyong.park@apctp.org},  ~~
Yun-Long Zhang$^{a,b,}$\footnote{e-mail : yunlong.zhang@apctp.org}\\
~\\
{\normalsize \it $^a\,$ Center for Quantum Spacetime and Department of Physics, Sogang University, Seoul 121-742, Korea} \\
{\normalsize \it $^b\,$ Asia Pacific Center for Theoretical Physics, Pohang, 790-784, Korea } \\
{\normalsize \it $^c\,$ Department of Physics, POSTECH, Pohang, 790-784, Korea }\\
}

\date{\normalsize (\today)}

\maketitle

\vskip2cm

\begin{abstract}\normalsize
We construct $SL(2,R)$ invariant in anisotropic medium, with a dual anisotropic charged black hole geometry in  massive gravity.
We show how to obtain $SL(2,R)$ elements in terms of new degrees of freedom for Electromagnetic configuration,
and construct the general expressions for conductivity with $SL(2,R)$ invariant.
The holographic conductivities can be calculated using horizon data in an external magnetic field,
and we show the numerical results using the linear response theory.
\end{abstract}


\vspace{1cm}

\vspace{2cm}


\end{titlepage}

\renewcommand{\thefootnote}{\arabic{footnote}}
\setcounter{footnote}{0}
%
%
%
\tableofcontents
\allowdisplaybreaks
\section{Introduction}
Gauge/gravity duality has become a 
well-established method to make a study on strongly interacting systems in (non-)conformal field theory. 
Motivated by M/Sting theory, Anti de sitter/Conformal Field Theory(AdS/CFT) correspondence is obtained from the 
low energy limit when field theory on brane decouples from the bulk. 
On such energy scale, AdS spacetime geometry can be matched with one-dimensional lower quantum field theory.  
AdS/CFT correspondence suggests that information about dual operator of large N coupling gauge field can be obtained from classical limit of supergravity as long as AdS radius is large. 
In general,  
the nonperturbative properties of strongly interacting system can be described by the dynamic of AdS geometry of classical gravity defined at  asymptotic AdS region \cite{Maldacena:1997re,Witten:1998qj, Witten:1998zw, Gubser:1998bc}. One of the main applications of AdS/CFT correspondence is the study of hydrodynamic, which is first investigated in \cite{Policastro:2002se, Policastro:2002tn}. 
This prescription opens a possibility to consider condense matter physics in strongly interacting many-body problem in point of view of holography. 
As a result, Anti de sitter/Condense Matter Theory correspondence has been intensively studied to construct a theoretical framework to describe, for example, the properties of transport coefficients.

 One of the surprising results predicted by the framework of holography related to conductivity is the quantum critical phenomena of cuprates. 
There exist the intermediate scaling behavior, and has been studied in different models such as lattice physics and massive gravity in isotropic structure \cite{Horowitz:2012ky, Horowitz:2012gs,Vegh:2013sk,Andrade:2013gsa,Davison:2013jba, Ling:2013nxa}. 
It is also generalised to the phenomena in anisotropic medium in \cite{Khimphun:2016ikw}. 
However, there are still some unsolved issues related to   intermediate scaling behavior (\ref{PowerLaws}) where the offset $b\neq 0$,
and we will discuss about how our present model motivated from spatially anisotropic system in \cite{Khimphun:2016ikw}, has improved the scenario. See also the study in anisotropic phase \cite{Jain:2015txa, Jain:2014vka}. It is also well-known that to consider transport coefficients in presence of translation symmetry will not lead to a well-defined DC conductivity. Thus, impurity should be introduced into the system which mimics the structure of real materials. To achieve this, people have considered transport function with a broken translation symmetry. To associate the system with momentum relaxation, people have considered models such as inhomogeneous scalar field \cite{Taylor:2008tg, Blake:2014yla, Kim:2015wba, Khimphun:2016ikw}, Q-lattice \cite{Chesler:2013qla, Donos:2014cya, Donos:2014yya, Ling:2013nxa} and non-linear massive gravity \cite{Vegh:2013sk, Andrade:2013gsa, Davison:2013jba, Zhou:2015dha}. In particular, momentum dissipation is turned on in two directions ($x$- and $y$-directions) by the axion fields leading to interesting results for hydrodynamic quantities \cite{ Khimphun:2016ikw}.
 
 However, in order to consider anisotropy in both directions associated with Electromagnetic (EM) duality, we can no longer treat axion field depending on linearly spatial coordinates which is due to inhomogeneity. As a result, we study a similar model based on \cite{Khimphun:2016ikw} with non-linear massive gravity due to the fact that conductivity in massive gravity is equivalent to spatial coordinate dependent scalar field model \cite{Andrade:2013gsa}. However, one will see that our present model with anisotropic massive gravity provides interesting results, particularly, the universal scaling behavior in an intermediate frequency regime. The gravity model we will consider includes a local gauge, dilaton, and axion fields, and all of  these are radial coordinate dependent. The massive term contains reference metric with an ansatz depending on radial coordinate as well which plays an important role in engineering anisotropic black brane. In order to check the properties of anisotropic medium, we study electric conductivity by turning on vector fluctuations. We compare AC conductivity near zero frequency limit with DC conductivity expressions obtained near horizon limit. After obtaining conductivity from electrically charged black brane, we discuss Dude form and intermediate scaling behavior. Then, from conductivities of the electrically charged configuration, we calculate Hall conductivity by using $SL(2,R)$ transformation. We will also discuss the cyclotron poles and compare our results with magnetohydrodynamic (MHD) studied in \cite{Hartnoll:2007ih, Hartnoll:2007ip} at appropriate limit where our system is approaching isotropy and translation symmetry. 
 
The rest of this paper is organized as follows. In {Secion}~\ref{Sec:AnisoBackground} we construct $SL(2,R)$ invariant dual charged black hole geometry with anisotropic medium and show how to obtain $SL(2,R)$ elements in terms of new degrees of freedom for EM configuration. We then calculate conductivity using horizon data in an external magnetic field and construct general expressions for conductivity with $SL(2,R)$ invariant in {Secion}~\ref{Sec:Conductivity}. After these set up, we show how anisotropic RN-AdS background solutions can be obtained from massive gravity model which will be used to consider linear response theory in {Secion}~\ref{Sec:ConductivityAdS}. We remark our results in {Secion}~\ref{Sec:Conclusion}.
\section{AdS Einstein-Maxwell-Dilaton-Axion Model in Massive Gravity}\label{Sec:AnisoBackground}
Recently, it has been well known that a finite DC conductivity requires breaking of the spatial translation symmetry. In the holographic model, there are several ways to break such symmetry.
One is to take into account a spatially linear axion field which leads to a momentum relaxation geometry \cite{Khimphun:2016ikw}. Another way is to introduce a graviton mass to break the diffeomorphism invariance. For the latter case, it was known that adding an additional Axionic-Chern-Simons term, $\tilde{a}\tilde{F} F$, gives rise to the electromagnetic (EM) duality at the equation of motion level. Since this Axionic-Chern-Simons term leads to a nontrivial Hall conductivity, the following gravity theory is useful to study the finite DC conductivity as well as the Hall conductivity of the dual field theory
\begin{align}\label{Action}
S= \int d^4 x \sqrt{- g}\Big(R +\dfrac{6}{L^2}-2 (\nabla\phi)^2-\dfrac{1}{2} e^{4\phi}(\nabla \tilde{a})^2 -e^{-2\phi} F^2-\tilde{a} F\tilde{F}+p_1[\mathcal{K}]+p_2([\mathcal{K}]^2-[\mathcal{K}^2])
\Big)\, ,
\end{align}
where $p_1$ and $p_2$ are constant parameters. $[\mathcal{K}]$ is the trace of square root tensor defined by $\mathcal{K}^\mu\,_\sm\mathcal{K}^\sm\,_\nu\equiv  g^{\mu\sm}f_{\sm\nu}$ \cite{deRham:2010kj,Hinterbichler:2011tt,Cao:2015cti} and $f_{\mu\nu}$ is the reference metric which breaks diffeomorphism in $x$- and $y$- directions
\begin{align}\label{RefMetric}
f_{\mu\nu}=
\text{diag} \left[
0,  0, k_1^2 H(z)^2, k_2^2 H(z)^2  
\right]\,.
\end{align}
Note that different values of $k_1$ and $k_2$ lead to an anisotropic geometry.
Einstein equations are obtained as
\begin{align}
R_{\mu\nu} - \frac{1}{2} R g_{\mu\nu}+ \frac{3}{L^2} g_{\mu\nu} = T_{\mu\nu} ,
\end{align}
with energy-momentum tensor
\begin{align}\label{EnerMomenTensor}
T_{\mu\nu} = &~ 2\, \nabla_\mu\phi\nabla_\nu\phi +\dfrac{1}{2}e^{4\phi}( \nabla_\mu \tilde{a}\nabla_\nu \tilde{a})
 +2 e^{-2\phi} {F_{\mu \rho}} {F_{\nu}}^{\rho}\nonumber\\
 &-\dfrac{1}{2} g_{\mu\nu}\Big(2(\nabla\phi)^2+\dfrac{1}{2} e^{4\phi}(\nabla \tilde{a})^2
 +e^{-2\phi}F^2-p([\mathcal{K}]^2-[\mathcal{K}^2]) \Big)\nonumber\\
 &-\dfrac{1}{2} p_1 \mathcal{K}_{\mu\nu}-p_2\left([\mathcal{K}]\mathcal{K}_{\mu\nu}-[\mathcal{K}^2]_{\mu\nu}\right)   \,.
\end{align}

Then, the equations of motion for all fields are summarized as 
\begin{align}
& R_{\mu\nu}= -\dfrac{3}{L^2}g_{\mu\nu}+2 \nabla_\mu\phi\nabla_\nu\phi+\dfrac{1}{2}e^{4\phi}( \nabla_\mu \tilde{a}\nabla_\nu \tilde{a})+2 e^{-2 \phi}F_{\mu\rho}F_\nu\,^\rho-\dfrac{1}{2}g_{\mu\nu}e^{-2\phi}F^2\nonumber\\
&\qquad\quad+\dfrac{1}{2} p_1\mathcal{K}_{\mu\nu}-p_2\left([\mathcal{K}] \mathcal{K}_{\mu\nu}-[\mathcal{K}^2]_{\mu\nu}    \right)\,,\label{Eq:Einstein}   \\
&\nabla_{\mu}(e^{-2\phi}F^{\mu\nu}+\tilde{a} \tilde{F}^{\mu\nu})=0\,,\label{Eq:Maxwell}\\
&\square \phi -\frac{1}{2} e^{4\phi}(\nabla \tilde{a})^2+\dfrac{1}{2}e^{-2\phi} F^2 
=0\,, \label{Eq:Dilaton}\\
&\square \tilde{a}+4\nabla_{\mu}\phi\nabla^{\mu} \tilde{a} -F_{\mu\nu}\tilde{F}^{\mu\nu}=0\,.\label{Eq:Axion}
\end{align}
Notice that there is no  $\tilde{F}$ term in Einstein equation, due to its definition which includes $1/\sqrt{-g}$. Also, in purely electric or magnetic  charge, either $F$ or $\tilde{F}$ automatically vanishes.
But we can study conductivity for EM field using $SL(2,R)$ invariance.
Introducing new complex variables to check the invariance under $SL(2,R)$ transformation
\begin{align}
\lambda=\lambda_1+i \lambda_2\equiv\tilde{a}+i e^{-2\phi} \,,           \qquad F_{\pm}=F\pm i \tilde{F}\,,
\end{align}
one can rewrite equation of motion (\ref{Eq:Einstein})-(\ref{Eq:Axion}) following the notation in \cite{Shapere:1991ta}
\begin{align}
&R_{\mu\nu}=  -\dfrac{3}{L^2}g_{\mu\nu}+\dfrac{1}{4\ld_2^2}\left(\nb_\mu\bar{\ld}\nb_\nu\ld+\nb_\nu\bar{\ld}\nb_\mu\ld   \right) +2 \ld_2  F_{\mu\sm} F_\nu\,^\sm-\dfrac{1}{2} \ld_2 g_{\mu\nu} F^2 \nonumber\\
&\qquad\quad+\dfrac{1}{2} p_1\mathcal{K}_{\mu\nu}-p_2\left([\mathcal{K}] \mathcal{K}_{\mu\nu}-[\mathcal{K}^2]_{\mu\nu} \right)\,,\label{Eq:EinsteinNew}\\
&\nb_\mu(\ld  F_{+}^{\mu\nu}-\bar{\ld}  F_{-}^{\mu\nu})=0\,, \label{Eq:MaxwellNew}\\
& 2  \ld_2 {\nb_\mu\nb^\mu\lambda} +2 i {(\nb_\mu\ld)(\nb^\mu\ld)} - i \ld_2^3 F_{-}^2    =0\,, 
\label{Eq:AxionDilatonNew}
\end{align}
where $\bar{\ld}$ is complex conjugate of $\ld$. Equations (\ref{Eq:EinsteinNew})-(\ref{Eq:AxionDilatonNew}) are invariant under $SL(2,R)$ transformation 
\begin{align}\label{DualTran}
\ld\rightarrow \dfrac{a\ld+b}{c \ld+d}\,, \qquad ad-bc=1 , \nonumber\\
F_{\mu\nu}\rightarrow (c\ld_1+d) F_{\mu\nu}-c \lambda_2 \tilde{F}_{\mu\nu} .
\end{align}
We can check this $SL(2,R)$ symmetry in a simpler way. First, it is invariant under shift symmetry of hermitian matrice 
\begin{align}\label{Tdual}
 \left(
\begin{array}{cc}
1  & b  \\
 0 & 1 
\end{array}
\right)\,,
\end{align}
so that $\ld\rightarrow\ld+b$ and it is trivial to see that equations of motion (\ref{Eq:EinsteinNew})-(\ref{Eq:AxionDilatonNew}) are invariant. The second one is the transformation under traceless unitary matrice 
\begin{align}\label{Sdual}
 \left(
\begin{array}{cc}
0  & 1  \\
 -1 & 0 
\end{array}
\right)\,,
\end{align}
so that $\ld\rightarrow -1/\lambda$ and $\bar{\ld}\rightarrow -1/\bar{\lambda}$, and the field strength transforms as 
\begin{align}\label{Eq:FieldStrenthTransform}
F_{+}\rightarrow -\ld F_{+}\,, \qquad
F_{-}\rightarrow -\bar{\ld} F_{-}\,.
\end{align}
Equation (\ref{Eq:MaxwellNew}) interchange with Bianchi identity\footnote{$\lambda F_+^{\mu\nu} - \bar{\lambda } F_-^{\mu\nu} \rightarrow F_+^{\mu\nu} -F_-^{\mu\nu}$. Now in terms of $F_{\pm}$, Bianchi identity for $F^{\mu\nu}$ takes the form  $\nabla_\mu \left( F_+^{\mu\nu} -F_-^{\mu\nu} \right)=2 i\nabla_\mu \tilde{F}^{\mu\nu}=0$.} and (\ref{Eq:AxionDilatonNew}) are  invariant  under (\ref{Eq:FieldStrenthTransform}).Furthermore, Einstein Equation (\ref{Eq:EinsteinNew}) is also invariant following the prescription in \cite{Sen:1992fr}.


Using the following metric ansatz
\begin{align}\label{Metric}
ds^2= \dfrac{L^2}{z^2}\left(-g(z)dt^2 +g(z)^{-1}dz^2+e^{2U_1(z)}dx^2+e^{2U_2(z)}dy^2\right)\,,
\end{align}
the last two terms of (\ref{EnerMomenTensor}) reduce to
\begin{align}\label{TraceMassTerm}
p_1 \mathcal{K}_{xx}+p_2\left([\mathcal{K}]\mathcal{K}_{xx}-(\mathcal{K}^2)_{xx}\right)= \frac{e^{U_1-U_2}}{4 z} \left(p_1 L k_2  e^{U_1}+3 p_1 L k_1  e^{U_2}+4 p_2 k_1 k_2 z\right)\,,\nonumber\\
p_1 \mathcal{K}_{yy}+p_2\left([\mathcal{K}]\mathcal{K}_{yy}-(\mathcal{K}^2)_{yy}\right)= \frac{e^{U_2-U_1}}{4 z} \left(p_1 L k_1  e^{U_2}+3 p_1 L  k_2  e^{U_1}+4 p_2 k_1 k_2 z\right)\,,
\end{align}
which implies that $T_{xx}\neq T_{yy}$ if $U_1\neq U_2$. Thus, the resulting solution must be anisotropic for $k_1 \ne k_2$, as mentioned before.
When we take the following background gauge field
\begin{align}\label{BackGuaget}
A_\mu dx^\mu= A_t(z) dt\,,
\end{align}
we can get the conserved electric charge, $Q= - e^{U_1+U_2-2\phi}A_t'$, from (\ref{Eq:Maxwell}). Since $\tilde{F}$ term automatically vanishes in this case, the resulting field strength becomes
\begin{align}\label{F}
F=  e^{-U_1-U_2}(\ld_2)^{-1}Q \,dt \wedge dz  \,.
\end{align}
Due to the coupling to the scalar field, this field strength induces an effective potential for the scalar field 
\begin{align}\label{EP}
V_{eff}=\ld_2^{-1} Q^2\,.
\end{align}

Similarly, if both electric and magnetic charges are allowed, the field strength satisfying the Maxwell equation becomes \cite{Goldstein:2010aw}
\begin{align}
\bar{F}= e^{-U_1-U_2}(\bar{\ld}_2)^{-1}(\bar{Q}_e-\bar{\ld}_1\bar{Q}_m) \,dt \wedge dz +\bar{Q}_m dx \wedge dy  \,.\label{Fbar}
\end{align}
Notice that we distinguish above EM charged excitation by using the bar-symbol to denote the new configuration after duality transformation 
 Notice that (\ref{Fbar}) satisfies (\ref{Eq:MaxwellNew}) as well as Bianchi identities $\nabla_{(\mu}\bar{F}_{\nu\rho)}=0$. Then, we also have scalar effective potential as been done in \cite{Iizuka:2007zz}
\begin{align}
\bar{V}_{eff}= \bar{\ld}_2^{-1}\left( \bar{Q}_e-\bar{\ld_1}\bar{Q}_m\right)^2 +\bar{\ld_2}\bar{Q}_m^2\,,\label{EP-EB}
\end{align}
and (\ref{EP-EB}) reduces to (\ref{EP}) for electrically charged solution with $\bar{Q}_m=0$. Starting from this electrically charged solution, after the EM duality transformation, we obtain 
\begin{align}
\bar{F}_{xy}&=\bar{Q}_m=\left(c \ld_1+d\right) F_{xy}- c\ld_2 \tilde{F}_{xy}=- c \ld_2 \tilde{F}_{xy}=c Q\,,\label{FxyBar}\\
\bar{F}_{tz}  &=\left( c\ld_1+d\right)F_{tz}- c\ld_2 \tilde{F}_{tz}=\left( c\ld_1+d\right)F_{tz},\label{FtrBar}
\end{align}
where $\tilde{F}_{xy}=-g_{xx}g_{yy}\epsilon^{xytz}F_{tz}=-(g_{xx}g_{yy}/\sqrt{-g})(e^{-(U_1+U_2)}Q/\ld_2)=-Q/\ld_2$ and $\tilde{F}_{tz}=0$. Comparing $\bar{F}_{tz}$ in (\ref{Fbar}) with (\ref{FtrBar}), we find the following relation
\begin{align}
\bar{Q}_e= \left( \dfrac{\bar{\ld_2}}{\ld_2} (c\ld_1+d)+c \bar{\ld_1}\right)Q\,.\label{QeBar}
\end{align}
Using $SL(2,R)$ transformation in (\ref{DualTran}), this relation can be further reduced to
\begin{align}
\bar{Q}_e=a\, Q\,.
\end{align}
Since an effective potential of the scalar field must be invariant under duality transformation, (\ref{EP}) and (\ref{EP-EB}) should be equal. This fact   gives rise to
\begin{align}
\ld_2=\dfrac{Q^2 \bar{\ld}_2}{\left(\bar{Q}_e-\bar{\ld}_1\bar{Q}_m\right)^2 +(\bar{Q}_m \bar{\ld}_2)^2}\,,\label{Lambda2}
\end{align}
and substituting this relation into (\ref{QeBar}) leads to
\begin{align}
d=\dfrac{Q\left(\bar{Q}_e- c Q \bar{\ld}_1\right)}{\left(\bar{Q}_e-\bar{\ld}_1\bar{Q}_m\right)^2 +(\bar{Q}_m \bar{\ld}_2)^2}\,,
\end{align}
where $c=\bar{Q}_m/Q$ from (\ref{FxyBar}).  From the  constraint $ad-bc=1$, lastly, we have
\begin{align}
b=-\dfrac{\bar{Q}_e\ld_1}{Q}+\dfrac{Q\left(\bar{Q}_e\ld_1-\bar{Q}_m(\bar{\ld}_1^2+\bar{\ld}_2^2)\right)}{\bar{Q}_e^2-2\bar{Q}_e \bar{Q}_m \bar{\ld}_1+\bar{Q}_m^2(\bar{\ld}_1^2+\bar{\ld}_2^2)}\,.
\end{align}
Above electrically charged system can be classified by four free parameters, $M$, $Q$, and asymptotic values of $\ld_{10}=\tilde{a} |_{z\rightarrow 0}$ and $\ld_{20} = e^{-2\phi}|_{z\rightarrow0}$. After the duality transformation, since the duality transformation renders three more parameters, the resulting system  has totally seven parameters. However, a dyonic solution requires only five independent parameters such as $\bar{M}$, $\bar{Q}_e$, $\bar{Q}_m$, $\bar{\ld}_{10}$ and $\bar{\ld}_{20}$. This fact implies that there are two parameter redundancies. In order to get rid of such redundancies, we can choose specific values for $\ld_{10}$ and $Q$. We take $\ld_{10}=0$ and $Q$ will be determined in terms of $k_1$ and $k_2$ in later sections.\footnote{It is more natural to see why we treat this boundary field in this way, particularly, when we consider a specific case in AdS solution where this field turns out to be constant.}

\section{Holographic Conductivities} 
\label{Sec:Conductivity}
In this section, we will analyze the properties of holographic conductivities.

\subsection{Conductivity in External Magnetic Field}\label{ExMagnet}
To get more information about the transformation of conductivity between $x$- and $y$- directions,
we first analyze the DC conductivity near horizon. Turning on the background magnetic field with $A_y=B x$, gauge and metric fluctuations are given as
\begin{align}
A_\mu dx^\mu\rightarrow A_t(z) dt +\epsilon(A_x(z)-E_x t)dx+[B x+\epsilon(A_y(z)-E_y t)]dy\,,\label{PGauge}\\
g_{\mu\nu}dx^{\mu}dx^{\nu}\rightarrow \bar{g}_{\mu\nu}dx^{\mu}dx^{\nu} +\epsilon\dfrac{2L^2}{z^2}[\delta g_{tx^i}(z)dtdx^i+\delta g_{zx^i}(z) dzdx^i]\,.\label{PBulk}
\end{align}
where $\epsilon$ indicates a perturbation parameter and $\bar{g}_{\mu\nu}$ is background metric in (\ref{Metric}).
First, we begin with the electrically charged case with $B=0$. Combining the perturbed $ti$- and $ii$-components of Einstein equation, one can algebraically obtain 
\begin{align}
\delta g_{zi}=\dfrac{4 E_i z^3 A_t' e^{U_1+U_2-2\phi}}{k_i H g( p_1 e^{U_j}+2 p_2 k_j z H )}\,.
\end{align}
Notice that we set $L=1$ for simplicity. In addition, the $ti$-component of Einstein equation yields
\begin{align}
4 z^2 A_t' e^{-2\phi}A_i'+(\dfrac{k_i H  e^{-U_i}(p_1 +2 p_2 z k_j H e^{-U_j})}{z g} & +\dfrac{4 U_i'}{z} -2 U_i'(U_j'+ U_i')+2U_i'' )\delta g_{ti} \nonumber  \\
 &  +(U_j'-U_i'-\dfrac{2}{z} ) \delta g_{ti}' +\delta g_{ti}''
=0\,,
\end{align}
where indices $i\neq j$ and $i, j=(x,y)=(1,2)$. Near the horizon ($z\rightarrow 1$) the gauge field $A_y$ is well-defined if $A_i\sim -E_i/g+\mathcal{O}(1-z)$, so that we have
\begin{align}\label{gtiH}
 \delta g_{ti}\approx 
\dfrac{4 E_i A_t' e^{U_1+U_2-2\phi}}{k_i H ( p_1 e^{U_j}+2 p_2 k_j z H )}+\mathcal{O}(1-z)\,,
\end{align}
 which also implies that $\delta g_{ti}\sim g \delta g_{zi}+\mathcal{O}(1-z)$.
As a result, fluctuations near the horizon can be expanded into
\begin{align}\label{ExpandSolsHorizonDC}
&  \delta A_i' = - \dfrac{E_i}{g}+ \mathcal{O}(1-z)\,, \qquad \delta \tilde{a} = \mathcal{O}((1-z)^0)\,, \qquad
   \delta g_{ti}= g\, \delta g_{zi}+\mathcal{O}(1-z)\,.
\end{align}

Now, let us turn on a nonvanishing external magnetic field. At linear order, the conserved currents along $x$- and $y$-directions can be represented as
\begin{align}
J_x &=- g e^{-U_a-2 \phi } A_x'+ z^2 Q e^{-2U_1} \delta  g_{tx}- g z^2 B e^{-U_t-2 \phi } \delta g_{zy} \nonumber\\
J_y &= - g e^{U_a-2 \phi } A_y'+ z^2 Q e^{-2U_2} \delta  g_{ty}+ g z^2 B e^{-U_t-2 \phi } \delta g_{zx} 
\end{align} 
Massaging Einstein equations with the above perturbative solution in (\ref{ExpandSolsHorizonDC}), we can have the following relations near horizon
\begin{align}
-B A_t' \delta g _{ty} e^{2 (U_1- U_2- \phi) }+B^2 \delta g _{tx} e^{-2 U_2-2 \phi }-\frac{1}{4} k_1 p_1 H e^{U_1} \delta g _{tx}  -\frac{1}{2} k_1 k_2 p_2 H^2 e^{U_1-U_2}\delta g _{tx} &+B e^{2(U_1-U_2-\phi)} E_y\nonumber\\
 &=  -E_x A_t' e^{2 U_1-2 \phi }\,,
\end{align}
\begin{align}
 B \delta g_{tx}  e^{-2 \phi } A_t'+ B^2 \delta g_{ty} e^{-2 U_2-2 \phi }-\frac{1}{4} p_1 k_2 H e^{U_1}\delta g_{ty}-\dfrac{1}{2} p_2 H^2 k_1 k_2 e^{U_1-U_2}\delta g_{ty} & +e^{2(U_1-\phi)} A_t' E_y\nonumber\\ 
 & = B E_x e^{-2 \phi }\,.
\end{align}
From the Ohm's law
\begin{align}\label{OhmLaw}
 \left(
\begin{array}{c}
   J_x    \\
   J_y
\end{array}\right)=\left( \begin{array}{cc}
   \sm_{xx} & \sm_{xy}  \\
    \sm_{yx} & \sm_{yy}
\end{array}
\right)
\left( \begin{array}{c}
   F_{tx}   \\
    F_{ty}
\end{array}
\right)  \,,
\end{align}
the corresponding DC conductivity near horizon are
\begin{align}\label{SigmaXXMagP20}
\bar{\sm}_{xx}\Big |_{z\rightarrow 1}& =\frac{\tilde{p}_1 k_2  e^{U_t+U_2} \left(\tilde{p}_1 k_1  e^{ U_t+ U_2+2 \phi }-4 B^2 -4 Q^2 e^{4 \phi }\right)}{16 B^4 +4 B^2 e^{2 \phi } \left(4 Q^2 e^{2 \phi }-\tilde{p}_1 k_1  e^{ U_t+ \text{U2}}-\tilde{p}_1 k_2  e^{ U_t+U_1}\right)+\tilde{p}_1^2 k_1 k_2 e^{3 U_t+4 \phi }}\,,\\
\bar{\sm}_{yy}\Big |_{z\rightarrow 1}&=\frac{\tilde{p}_1 k_1 e^{U_t+ U_1} \left(\tilde{p}_1 k_2 e^{ U_t+ U_1+2 \phi }-4 B^2 -4 Q^2 e^{4 \phi }\right)}{16 B^4 +4 B^2 e^{2 \phi } \left(4 Q^2 e^{2 \phi }-\tilde{p}_1 k_1  e^{ U_t+ U_2}-\tilde{p}_1 k_2 e^{U_t+U_1}\right)+\tilde{p}_1^2 k_1 k_2 e^{3 U_t+4 \phi }}\label{SigmaYYMagP20}\,,\\
\bar{\sm}_{xy}\Big |_{z\rightarrow 1}&=\frac{4 B Q \left(\tilde{p}_1 k_1 e^{3 U_t+ U_2+2 \phi }+\tilde{p}_1 k_2 e^{3U_t+U_1+2 \phi }-4 B^2 e^{2 U_t}-4 Q^2 e^{4 \phi }\right)}{16 B^4 e^{3 U_t}+4 B^2 e^{U_t+2 \phi } \left(4 Q^2 e^{2 \phi }-\tilde{p}_1 k_1 e^{3 U_t+ U_2}-\tilde{p}_1 k_2  e^{3U_t+U_1}\right)+\tilde{p}_1^2 k_1 k_2 e^{6 U_t+4 \phi }}\,,
\end{align}
where we denote $U_t=U_1+U_2$, $U_a=U_1-U_2$, $Q=-e^{U_t-2\phi}A_t'$ and $ \tilde{p}_1=p_1 H$. Notice that we obtain above results without considering quadratic order of the massive term $\mathcal{O}([\mathcal{K}]^2)$, in other word, $p_2=0$. These results imply that $\bar{\sm}_{xx}= e^{- U_a} \bar{\sm}_{yy}$ and $\bar{\sigma}_{yx}=-\bar{\sigma}_{xy}$ under $(x,y)\rightarrow (e^{-U_a/2}y, -e^{U_a/2}x)$. This is consistent with the result obtained in \cite{Blake:2014yla} when $k_1=k_2$.
In this model, we can see that there exists an anomalous scaling behavior between the Hall angle and the DC conductivity. The similar behavior has also been noticed in the non-linear theory of the massive gravity for a Reissner-Nordstr\"{o}m-AdS and hyperscaling violation geometries \cite{Blake:2014yla,Zhou:2015dha}. 
In the next sections, we will check the above analytic results numerically motivated by membrane paradigm \cite{Parikh:1997ma, Iqbal:2008by} .


\subsection{Conductivity from $SL(2,R)$ invariance}
Applying the transformation, $(x,y)\rightarrow (G\, y, -G^{-1} x)$, used in the previous {Secion}~\ref{ExMagnet}, the conserved currents at the boundary ($z\rightarrow 0$) can be written as
\begin{align}
J_x\Big |_{z\rightarrow 0} &=4\left( \sm_{2} F_{tx}-\sm_{1} F_{ty}\right)=4\left(\ld_2 F_{zx}-\ld_{1} F_{ty}\right)\label{CurrentSigmax}\\
 J_y\Big |_{z\rightarrow 0} &=4\left( \sm_{1} F_{tx}+ G^{-2} \sm_{2} F_{ty}\right)=4\left(\ld_2 F_{zy}+\ld_{1} F{tx}\right)\label{CurrentSigmay} ,
\end{align}
where we define $\sm_1=\sm_{yx}/4$ and $\sm_2=\sm_{xx}/4$. This represents the linear response theory of the dual field theory. Now, let us study how the $SL(2,R)$ transformation modifies the above response theory. For later convenience, we introduce new variables
\begin{align}\label{SigmaPM}
\sm_{\pm}\equiv \sm_{1}\pm i G^{-2}\sm_2 \,.
\end{align}
Intriguinly, we find that the linear response theory is invariant under the $SL(2,R)$ transformation if the newly defined variables trnasform as
\begin{align}\label{SigmaPMTran}
\sm_\pm\rightarrow \dfrac{a\sm_\pm+b}{c \sm_\pm+d}\,.
\end{align}
To see in more details, let us investigate how (\ref{CurrentSigmax}) changes under the $SL(2,R)$ transformation.
Since the shift transformation in (\ref{Tdual}),  $\ld_1\rightarrow \ld_1+b$ and $\sm_1\rightarrow\sm_1 +b$, does not change (\ref{CurrentSigmax}), we focus on the transformation in (\ref{Sdual}). If (\ref{CurrentSigmax}) is still invariant under the transformation in (\ref{Sdual}), we can conclude that the response theory is really invarinat under the $SL(2,R)$ transformation in (\ref{DualTran}).

The right hand side of (\ref{CurrentSigmax}) can be rewritten as 
\begin{align}
\ld_2 F_{zx}-\ld_{1} F_{ty}\equiv -\dfrac{\ld}{2} (F_{+})_{ty}-\dfrac{\bar{\ld}}{2} (F_{-})_{ty}\,,
\end{align}
by using (\ref{Eq:FieldStrenthTransform})
\begin{align}\label{RHSTran}
-\dfrac{\ld}{2} (F_{+})_{ty}-\dfrac{\bar{\ld}}{2} (F_{-})_{ty}\rightarrow -F_{ty} .
\end{align}
On the other hand, the left hand side of (\ref{CurrentSigmax}) reduces to
\begin{align}\label{LHS}
\sm_2 F_{tx} -\sm_1 F_{ty}& \equiv -\dfrac{1}{2}\sm_{+}(F_{ty}+ i F_{tx})-\dfrac{1}{2}\sm_{-}(F_{ty}-i F_{tx})\nonumber\\
&\rightarrow  -\dfrac{1}{2} \left(-\dfrac{1}{\sm_{+}}\right)\left[\ld_2 \tilde{F}_{ty}-\ld_1 F_{ty}+ i (\ld_2 \tilde{F}_{tx}-\ld_1 F_{tx})\right]\nonumber\\
&\quad\,\,-\dfrac{1}{2}\left(-\dfrac{1}{\sm_{-}}\right)\left[ \ld_2 \tilde{F}_{ty}-\ld_1 F_{ty}-i(\ld_2 \tilde{F}_{tx}-\ld_1 F_{tx})  \right]\,.
\end{align}
Here, the transformation in (\ref{Sdual})  used in the second line where $\sm_{\pm}\rightarrow -1/\sm_{\pm}$ and $F_{\mu\nu}\rightarrow \ld_2 \tilde{F}_{\mu\nu} -\ld_1 F_{\mu\nu} $. With a bit of algebra, we obtain
\begin{align}
\sm_2 F_{tx} -\sm_1 F_{ty}\rightarrow \dfrac{1}{\sm_{+}\sm_{-}} \left[ \sm_1 (\ld_2 F_{zx}-\ld_1 F_{ty})- G^{-2} \sm_2 (\ld_2 F_{zy}+\ld_1 F_{tx})\right]\,.
\end{align}
By using the right hand side of (\ref{CurrentSigmax}), we have 
\begin{align}\label{LHSTran}
\sm_2 F_{tx} -\sm_1 F_{ty}\rightarrow - F_{ty}\,.
\end{align}
According to (\ref{RHSTran}) and (\ref{LHSTran}), we can show that (\ref{CurrentSigmax}) is invariant under $SL(2,R)$ transformation. The same is true for (\ref{CurrentSigmay}).

Using the fact that $\sm_1=\sm_{yx}/4=0$ for an electrically charged system, old and new conductivities after the $SL(2,R)$ transformation are related to each other
\begin{align}   \label{SigmaBarXYGeneral} 
\bar{\sm}_{xx}&= \dfrac{\sm_{xx}}{d^2+c^2 G^{-4}(\sm_{xx}/4)^2}= \dfrac{\sm_{xx}}{d^2+c^2 (\sm_{yy}/4)^2}\,,\nonumber\\
\bar{\sm}_{yx}&= 4\dfrac{a c\, G^{-4} (\sm_{xx}/4)^2+bd}{d^2+c^2 G^{-4}(\sm_{xx}/4)^2}= 4\dfrac{a c\, (\sm_{yy}/4)^2+bd}{d^2+c^2 (\sm_{yy}/4)^2}  \,, \nonumber\\
\bar{\sm}_{yy}&= \dfrac{G^{-2}\sm_{xx}}{d^2+c^2 G^{-4}(\sm_{xx}/4)^2}= \dfrac{\sm_{yy}}{d^2+c^2 (\sm_{yy}/4)^2} \,,\nonumber\\
\bar{\sm}_{xy} &= -4\dfrac{a c\, G^{-4} (\sm_{xx}/4)^2+bd}{d^2+c^2 G^{-4}(\sm_{xx}/4)^2}= -4\dfrac{a c\,(\sm_{yy}/4)^2+bd}{d^2+c^2 (\sm_{yy}/4)^2}  \,,
\end{align}
where $G^{-2}=\sm_{yy}/\sm_{xx}$. This result shows that after transformation, $\bar{\sm}_{yy}=G^{-2}\bar{\sm}_{xx}$ and $\bar{\sm}_{xy}=-\bar{\sm}_{yx}$.  

\section{Numerical Calculations}
\label{Sec:ConductivityAdS}
\begin{figure}
\begin{center}
\includegraphics[height = 0.16\textheight]{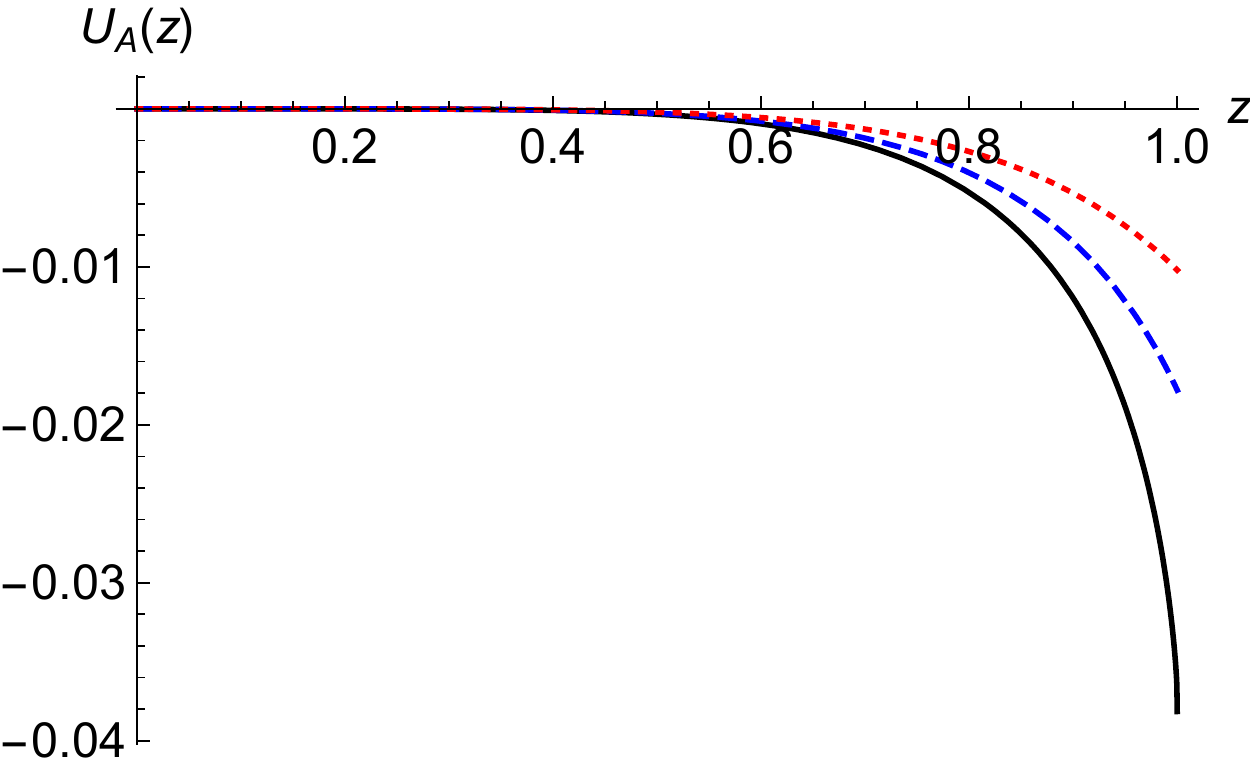}~~
\includegraphics[height = 0.16\textheight]{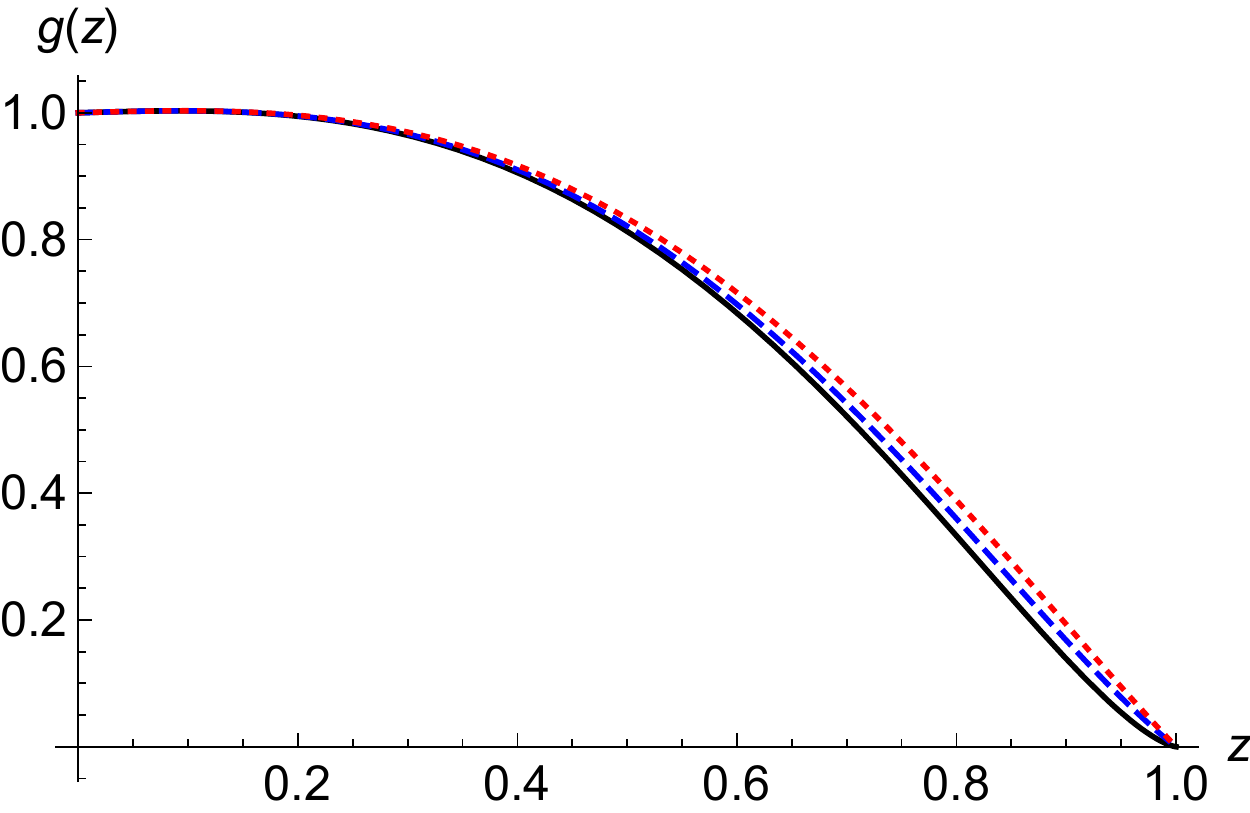}~~
\includegraphics[height = 0.16\textheight]{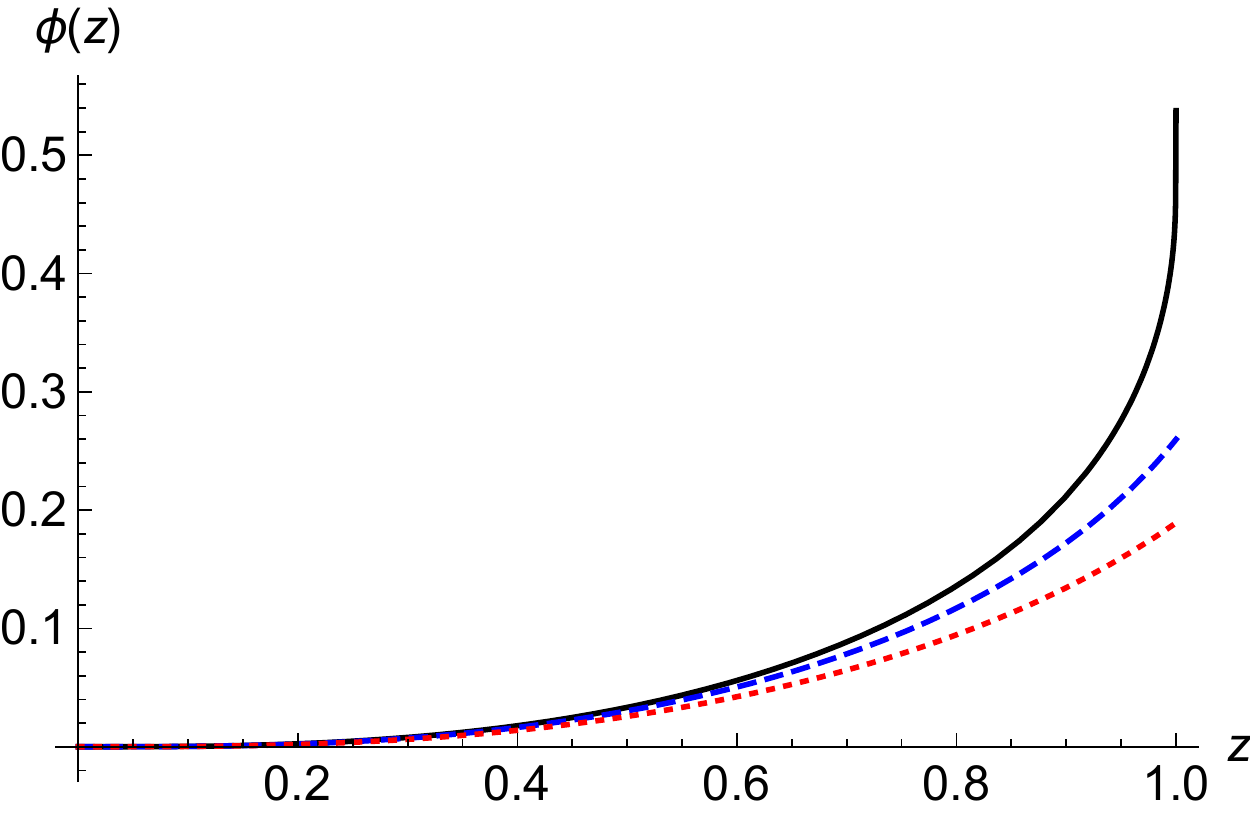}
\caption{Plots of the functions (Isotropic case) in terms of $z$, for fixed 
$H(z)=1$, $p_1=1$, $p_2=1$, $k_1=k_2=0.1$, with $\kappa=0.001$ (Black curve), $\kappa=1.35$ (Blue-dashed curve), and $\kappa=1.81$ (Red-dotted curve).}\label{BackSolIso}
\end{center}
\end{figure}

\begin{figure}
\begin{center}~~~
\includegraphics[height = 0.23\textheight]{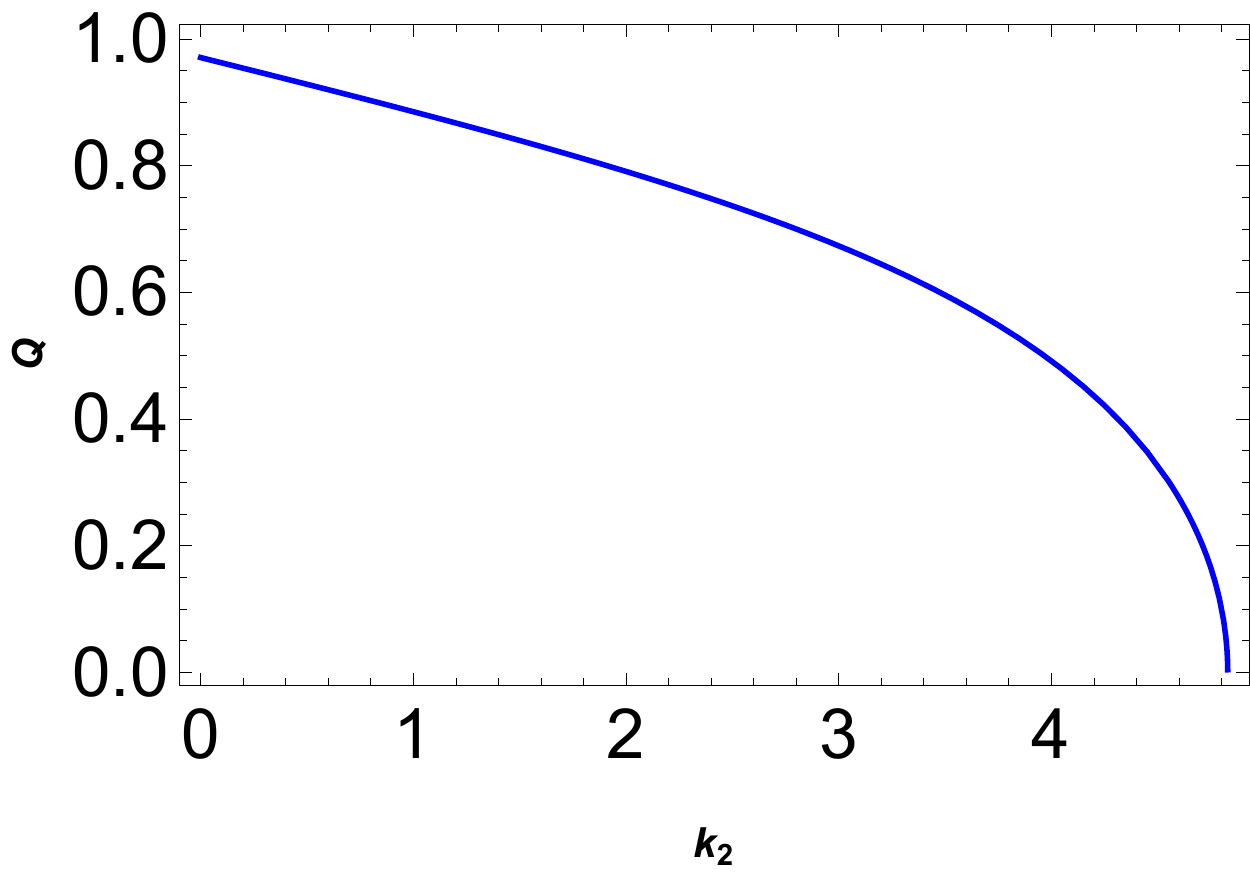}~
\includegraphics[height = 0.23\textheight]{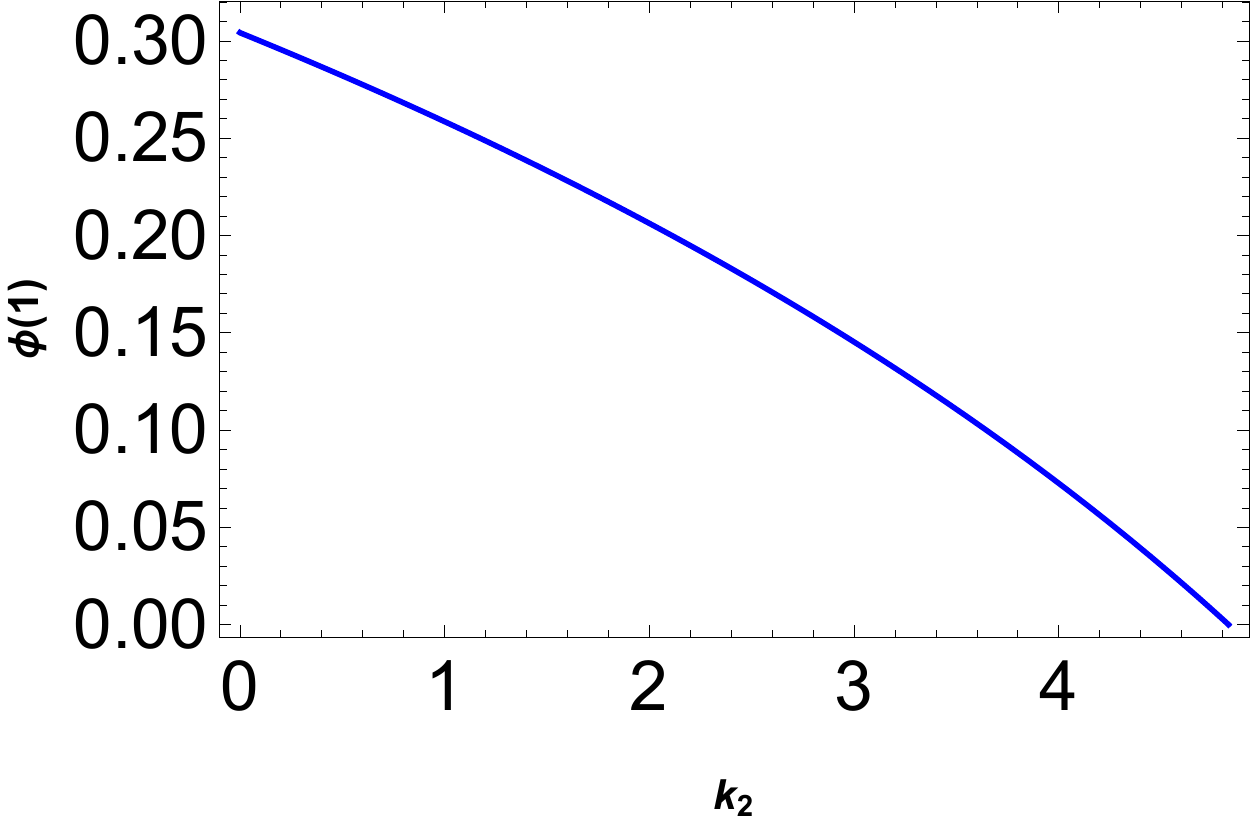}\\~~\\~~\\
\includegraphics[height = 0.23\textheight]{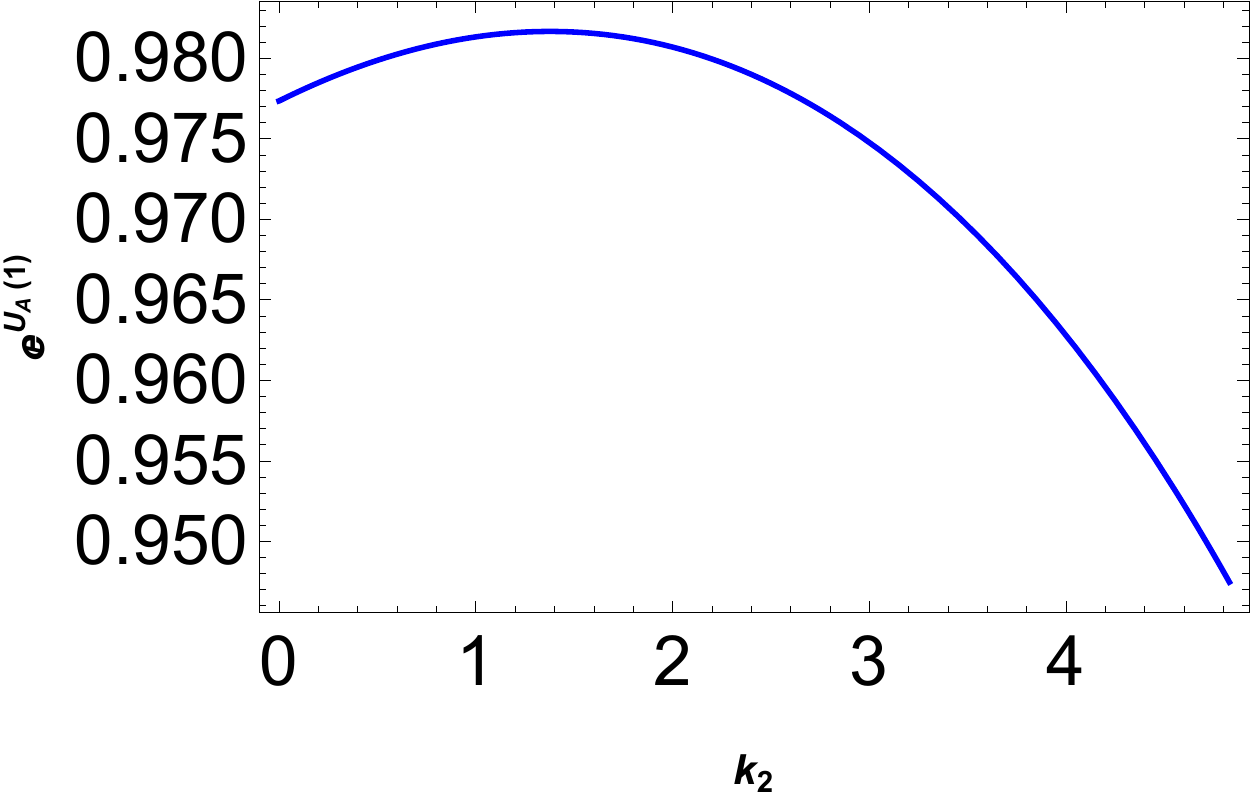}~
\includegraphics[height = 0.23\textheight]{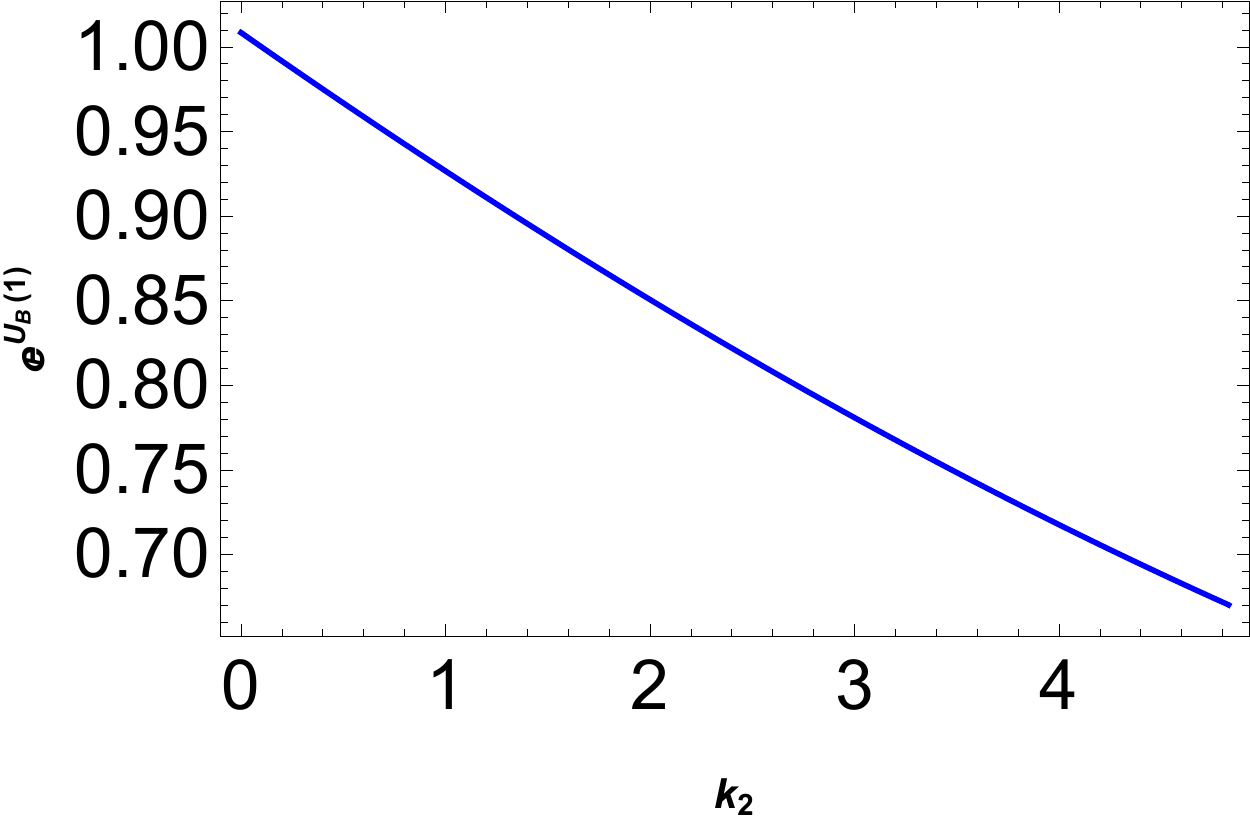}
\caption{Horizon values of the functions in terms of $k_2$, for fixed 
$p_1=-1$, $p_2=0$, $k_1=0.1$, $\kappa=1$.}\label{HorizonDataNum}
\end{center}
\end{figure}

\begin{figure}
\begin{center}
\includegraphics[height = 0.3\textheight]{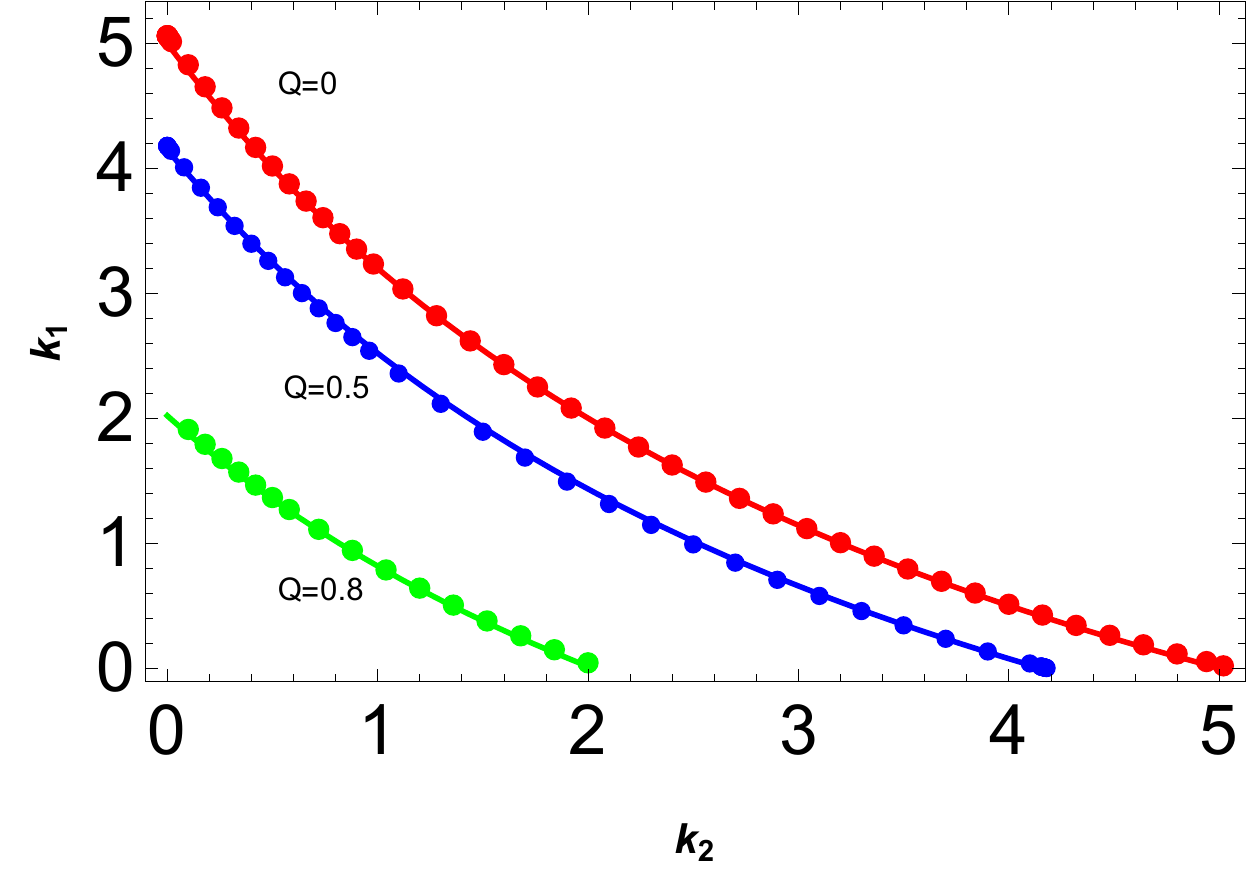}
\caption{ Fix $p_1=-1$, $p_2=0$, $\kappa=1$
Curves are from data fitting, and the fitting function matches well with $4(k_1+k_2)+k_1 k_2=20+6Q-26Q^2$. }\label{k1k2ConstQ}
\end{center}
\end{figure}

\begin{figure}
\begin{center}
\includegraphics[height = 0.2\textheight]{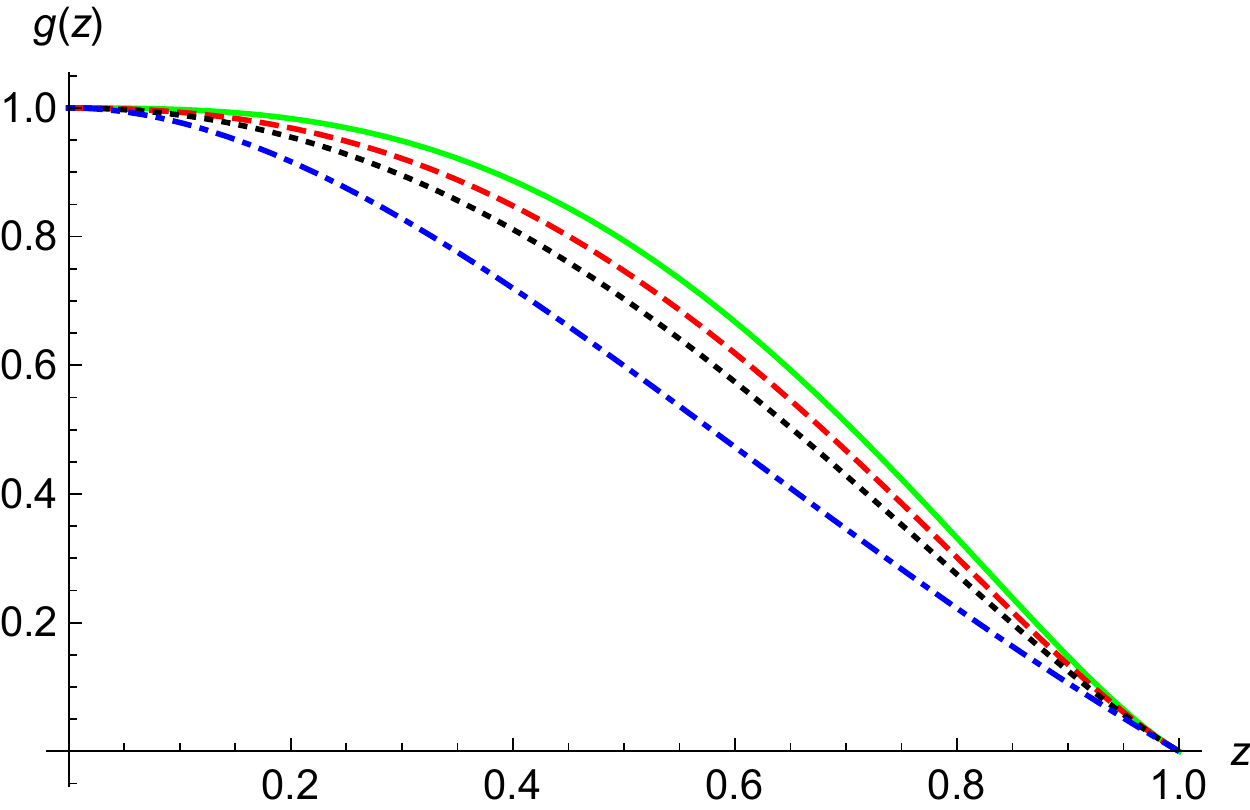}~~~~~
\includegraphics[height = 0.2\textheight]{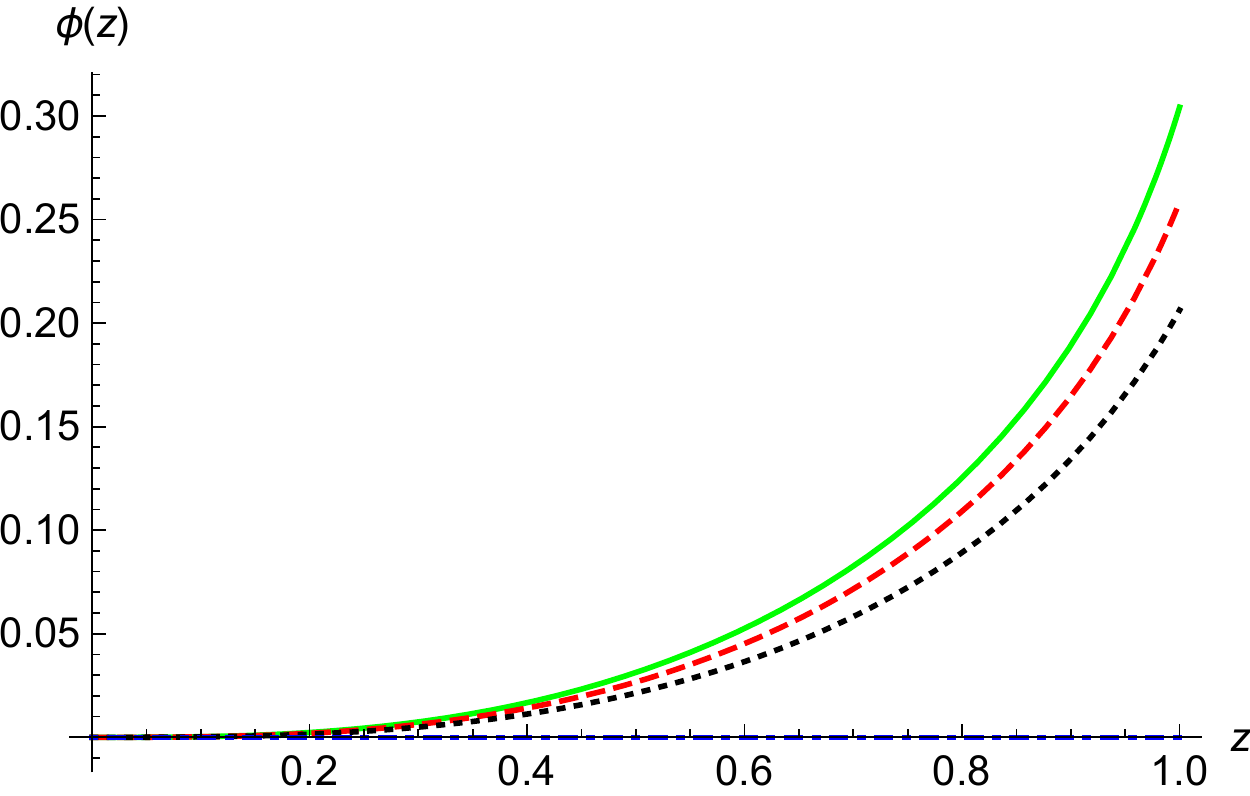}\\~~\\
\includegraphics[height = 0.2\textheight]{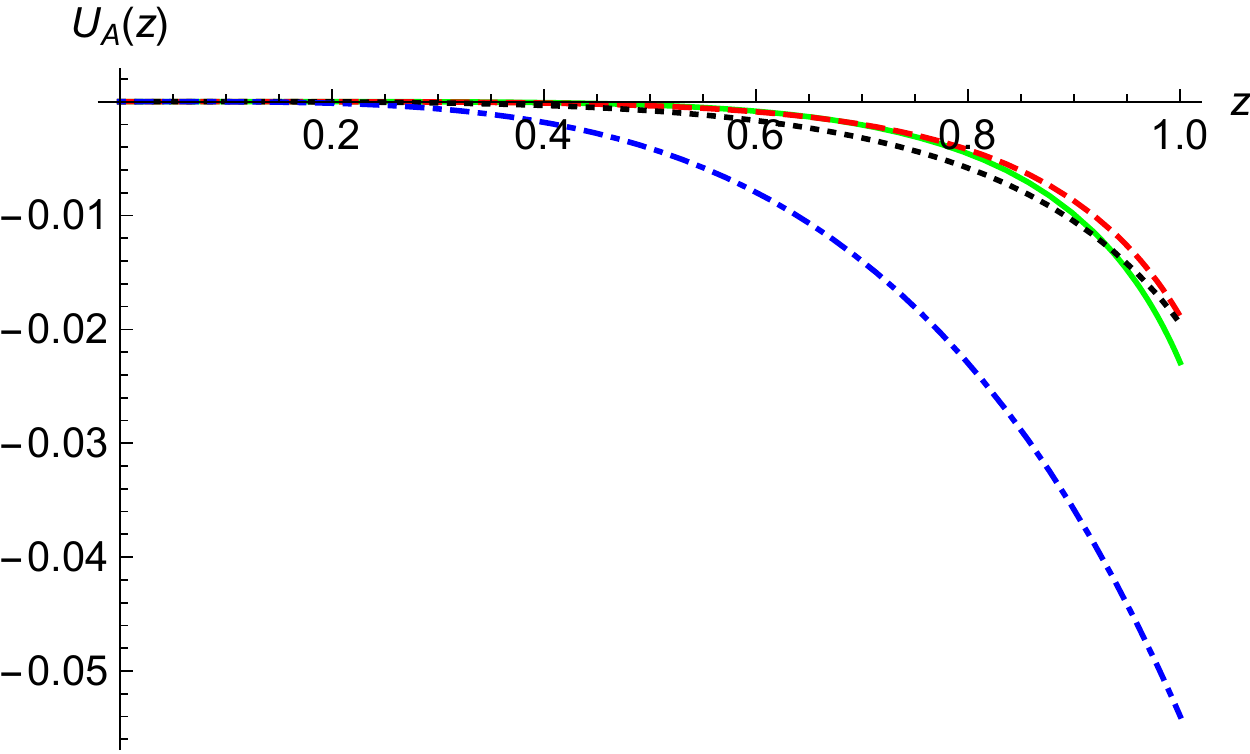}~~~
\includegraphics[height = 0.2\textheight]{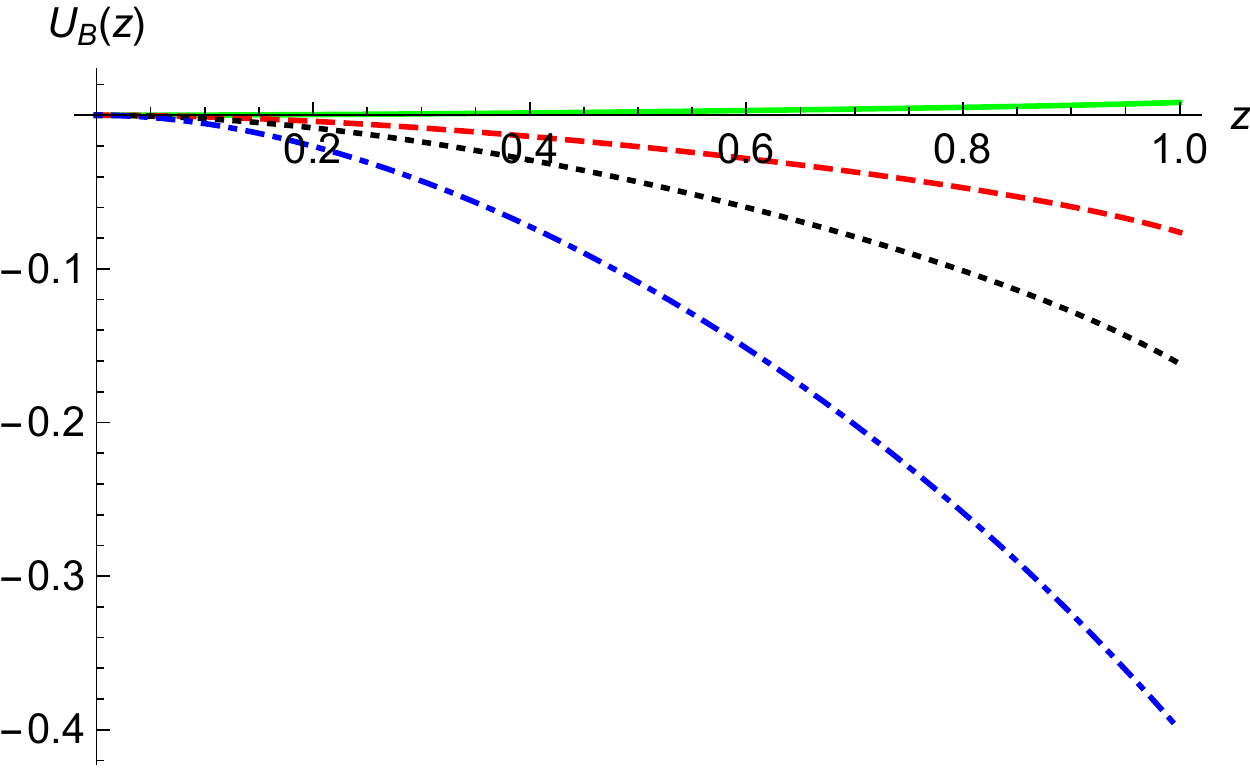}
\caption{Plots of the functions in terms of $z$, for fixed 
$p_1=-1$, $p_2=0$, $k_1=0.1$, $\kappa=1$,
with $k_2=0$ (Green), $k_2=1$ (Red-dashed), $k_2=2$ (Black-dotted), and $k_2=4.8294$ (Blue-dotted-dashed).  }\label{Fig:BackSolAniso}
\end{center}
\end{figure}

In this section, we will study the electric conductivity and the Hall conductivity caused by $SL(2,R)$ transformation. 
Considering the case where $p_2=0$,  the equations of motion are 
\begin{align}
\phi '^2+\frac{1}{4} e^{4 \phi } \tilde{a} '^2+U_A'^2+U_B'^2 +U_A''&= 0\,,\label{Eq:UA}\\
 \frac{1}{4} \tilde{p}_1 e^{-U_A-U_B} \left(k_2 e^{2 U_B}-k_1\right)+ \left(2 g \left(z U_A'-1\right)+z g'\right)U_B'+g z U_B'' &=0\,, \label{Eq:UB}\\
-4 z^4 Q^2 e^{2 \phi -4 U_A}+  \left(16 z U_A'-4 z^2 U_A'^2+4 z^2 U_B'^2+4 z^2 \phi '^2+z^2 e^{4 \phi } \tilde{a} '^2-12\right)g+12
\nonumber\\+2 \tilde{p}_1 e^{-U_A}( k_1 z^3 e^{-U_B}+2 k_2 z^3 e^{U_B})+4 k_1 k_2 p_2 H^2 z^4 e^{-2 U_A} + 4 z\left(1- z U_A'\right)g' &=0\,,\label{Eq:Constraint}\\
Q^2 z^3 e^{2 \phi -4 U_A}-\frac{1}{2} g z e^{4 \phi } \tilde{a} '^2
+ \left(2 g z U_A'+z g'-2 g\right)\phi '+g z \phi '' &=0\,,\\
 \left(2 g \left(z U_A'+2 z \phi '-1\right)+z g'\right)\tilde{a}'+g z \tilde{a} '' &=0 \label{Eq:Chi}\,,
\end{align}  
where we have defined $U_1=U_A+U_B$, $U_2=U_A-U_B$, and $Q=-e^{2U_A-2\phi}A_t'$.
Expanding variables near $z=0$ and substituting them into the above equations of motion, we find that if $\tilde{p}_1\equiv  p_1 H(z)$ does not vanish at the boundary, (\ref{Eq:UB}) cannot be satisfied for $k_1\neq k_2$. This implies that, for $H(z)=$constant, only the isotropic solution is allowed. In {Fig.}~\ref{BackSolIso}, we depict the numerically isotropic solution with $H(z)=1$, $k_1=k_2$ and $U_B(z)=0$. However, if $H(z)$ is given by a function of $z$ and suppressed rapidly at the boundary, an anisotropic solution as well as isotropic one are possible. From now on, we concentrate on an anisotropic black brane solution with $H(z)=z$. Note that equations of motion in (\ref{Eq:UA})-(\ref{Eq:Chi}) are invariant under the shift of scalar field 
\begin{align}
\phi\rightarrow \phi-\phi_0\,, \qquad Q\rightarrow e^{\phi_0} Q\,, \qquad  \tilde{a}\rightarrow e^{2\phi_0} \tilde{a},
\end{align}
and under the scaling of coordinates 
\begin{align}
&x \rightarrow e^{-(U_A(0)+U_B(0))}x\,, \qquad  y \rightarrow e^{-(U_A(0)-U_B(0))}y\,, \nonumber\\
&  k_1 \rightarrow e^{-(U_A(0)+U_B(0))}k_1\,,\qquad  k_2 \rightarrow e^{-(U_A(0)-U_B(0))}k_2\,.
\end{align}
The existence of the horizon requires $g(z)$ to be $g(1)=0$ at the horizon. Due to the above scaling symmetries, in addition, the other variables at the horizon can be fixed as
\begin{align}
&\chi'(1)=0\,, \qquad \phi'(1)=\dfrac{2(k_1+k_2+2k_1k_2)+Q^2}{\kappa}\,,\nonumber\\
& U_A'(1)= \dfrac{2(\kappa+Q^2)-(6+k_1+k_2+k_1k_2)}{2\kappa}\,, \qquad U_B'(1)= \dfrac{k_2-k_1}{4\kappa},
\end{align}
where $\kappa=-g'(1)$ is associated with the Hawking temperature via $\kappa=4 \pi T$.
In order to solve background equation of motion, we need horizon data as in {Fig.}~\ref{HorizonDataNum}. Plot of $Q$ shows that there exist a critical value at $k_2\approx 4.8294 $ where $Q$ approaches zero.
In order to see how such a critical point relies on the values of $k_1$ and $k_2$, we find numerically the critical point with various value of $k_1$ and $k_2$. Interestingly, {Fig.}~\ref{k1k2ConstQ} shows that numerical data are well fitted by $4(k_1+k_2)+k_1 k_2=20+6Q-26Q^2$. In {Fig.}~\ref{Fig:BackSolAniso}, we draw various bulk solutions with different values of $k_2$, which have an asymptotic AdS geometry. On these numerical background solution, hereafter we will investigate hydrodynamic quantities by perturbing vector and metric fluctuations.


\subsection{Fluctuation}
In order to obtain conductivity, we turn on the vector mode fluctuation along $x$ and $y$ components
\begin{align*}
A_a\rightarrow A_t(z)(dt)_a +\left[\tilde{A}_x(t,z)(dx)_a+\tilde{A}_y(t,z)(dy)_a \right]\,,
\end{align*}
and $ti$- and $zi$-components of metric fluctuation 
\begin{align*}
g_{ab}\rightarrow  \overline{g}_{ab}+2\dfrac{L^2}{z^2}\left[\delta\tilde{g}_{tx^i}(t,z)(dtdx^i)_{ab}+\delta\tilde{g}_{zx^i}(t,z)(dtdx^i)_{ab}\right]\,,
\end{align*}
where $\overline{g}_{ab}$ is the background metric in (\ref{Metric}). Using the Fourier mode expansion
\begin{align*}
\tilde{A}_i(t,z)= \int_{-\infty}^{\infty} \frac{d\Omega}{2\pi}e^{-i \Omega t}A_i(z),\quad
\delta\tilde{g}_{ti}(t,z)= \int_{-\infty}^{\infty} \frac{d\Omega}{2\pi} e^{-i \Omega t}\delta g_{ti}(z)\,, \quad \delta\tilde{g}_{zi}(t,z)= \int_{-\infty}^{\infty} \frac{d\Omega}{2\pi}i \Omega e^{-i \Omega t}\delta g_{zi}(z)\,.
\end{align*}
\eqref{Eq:UA}-\eqref{Eq:Chi} are reduced to

\begin{align}
\left(\dfrac{k_i p_1 Q H e^{-3 U_A-\epsilon_{i}U_B+2\phi}}{z} \right)\delta g_{zi}+ \left(\frac{\Omega ^2}{g^2}-\frac{4 Q^2 z^2 e^{-4 U_A+2 \phi }}{g}\right)A_i\nonumber\\
+ \left(-2\epsilon_{i} U_B'+\frac{g'}{g}-2 \phi '\right)A_i'+A_i'' &=0\,,\label{Peq:Ai}\\
-4 Q z^2  e^{-2 U_A} A_i-\left(\Omega^2+\dfrac{k_i p_1 H g e^{-U_A-\epsilon_{i}U_B}}{ z}   \right)+2\left(- U_A'-\epsilon_{i}U_B \right)\delta g_{ti}+\delta g_{ti}' &=0\,,\label{Peq:gti}\\
 \dfrac{\delta g_{ti}}{g^2}+\left( \dfrac{H'}{H}-\dfrac{3}{z} +\dfrac{g'}{g} +U_A-\epsilon_{ij}U_B\right)\delta g_{zi}+\delta g_{zi}'&=0\,, \label{Peq:gzi}
\end{align}
where $\epsilon_{x}=-\epsilon_{y}=1$. 

In order to solve above equations, let us first focus on the near horizon behavior.
 Equations (\ref{Peq:Ai})-(\ref{Peq:gzi}) are singular at the horizon due to $g(1)=0$. To satisfy equations, fluctuations must have appropriate singular behaviors at the horizon.  
Introducing new variables with an appropriate exponent $\gamma$,
\begin{align}
\hat{A}_i(z)\equiv g(z) A_i'(z)\,, 
\end{align}
and 
\begin{align}
A_i(z)&=(1-z)^\gamma a_i(z)\,, \qquad  ~ \hat{A}_i(z) =(1-z)^\gamma\hat{a}_i(z)\,,\nonumber \\
g_{ti}(z)& =(1-z)^\gamma\zeta_{ti}(z)\,,\quad ~ ~ g_{zi}(z)=(1-z)^\gamma\zeta_{zi}(z)/g(z).
\end{align}
the above equations reduce to four first-order differential equations near the horizon
\begin{align}
\hat{a}_i'- \left(2\epsilon_i U_B'+\frac{\gamma }{1-z}+2 \phi '\right)\hat{a}_i+ \left(\frac{\Omega ^2}{g}-4 z^2 e^{2 \phi -4 U_A} Q^2\right)a_i-\dfrac{1}{z}p_1 k_i H e^{2\phi-3U_A-\epsilon_i U_B} \zeta_{zi} &=0\,,\label{PHeq:aih}\\
a_i'-\frac{\gamma}{1-z}a_i-\frac{\hat{a}_i}{g} &=0\,,\label{PHeq:ai}\\
\zeta_{zi}'+ \left(\dfrac{H'}{H}+U_A'-\epsilon_iU_B'-\dfrac{\gamma}{1-z}-\dfrac{3}{z}\right)\zeta_{ti}+ \dfrac{\zeta_{ti}}{g} &=0\,\label{Peq:Zetati}\,,\\
\zeta_{ti}' -\left( 2\left(U_A'+\epsilon_i U_B'\right)+\dfrac{\gamma}{1-z}\right)\zeta_{ti}- \left( \dfrac{\Omega^2}{g}+\dfrac{1}{z}p_1 k_i H e^{-(U_A+\epsilon_i U_B)}\right) \zeta_{zi}-4 z^2 Q e^{-2 U_A} a_i&=0 .
\end{align}
Rewriting these equations as the eigenvalue equation
\begin{align}
 \left(
\begin{array}{cccc}
0  & -\frac{\Omega ^2}{g'(1)} & 0 & 0  \\
 \frac{1}{g'(1)} & 0  & 0 & 0  \\
 0 & 0 & 0  & -\frac{1}{g'(1)}  \\
 0 & 0 & \frac{\Omega^2}{g'(1)} & 0  \\
\end{array}
\right) \left(
\begin{array}{c}
 \hat{a}_i \\
 a_i \\
 \zeta_{zi}  \\
 \zeta_{ti} \\
\end{array}
\right)&=\gamma  \left(
\begin{array}{c}
 \hat{a}_i \\
 a_i \\
 \zeta_{zi}  \\
 \zeta_{ti} \\
\end{array}
\right)\,,\label{EigenEquationX}
\end{align}
it leads to the degenerated eigenvalues, $\gamma=\pm \frac{i \Omega}{g'(1)}$ and its eigenvectors are given by (see \cite{Khimphun:2016ikw,Koga:2014hwa} for more details) 
\begin{align}
\psi_{1\pm} =\left(
\begin{array}{c}
 0 \\
 0 \\
\pm i/ \Omega  \\
 1 \\
\end{array}
\right)\,,\quad
\psi_{2\pm} =\left(
\begin{array}{c}
\pm  i \Omega  \\
 1 \\
 0 \\
 0 \\
\end{array}
\right)\, . \label{EigenVectorH}
\end{align}

\subsection{Numerical Result}
\begin{figure}
\begin{center}
\includegraphics[height = 0.25\textheight]{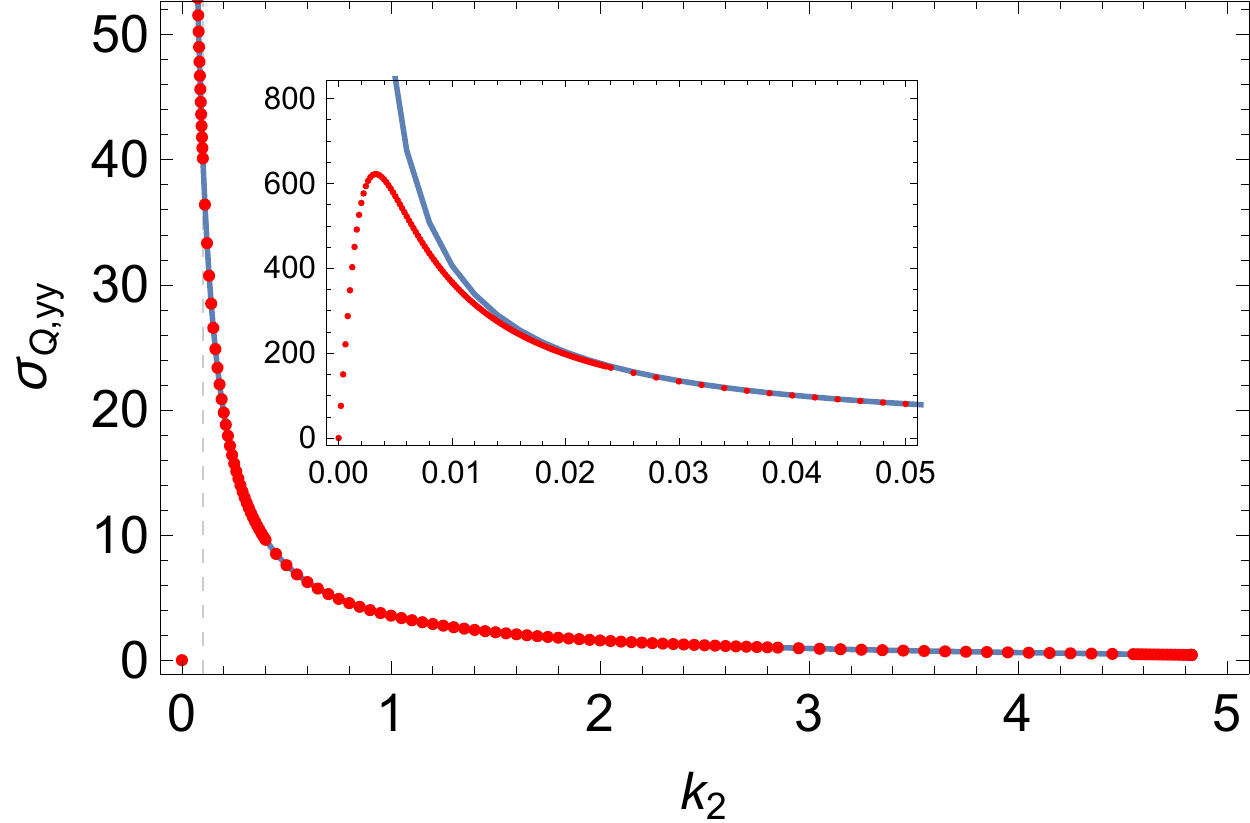}
\caption{$\sigma_{Q,yy}$ in terms of $k_2$, for fixed 
$p_1=-1$, $p_2=0$, $k_1=0.1$, and $k_2=0.4$.}\label{Fig:CompAnalytic}
\end{center}
\end{figure}
\begin{figure}
\begin{center}
\includegraphics[height = 0.25\textheight]{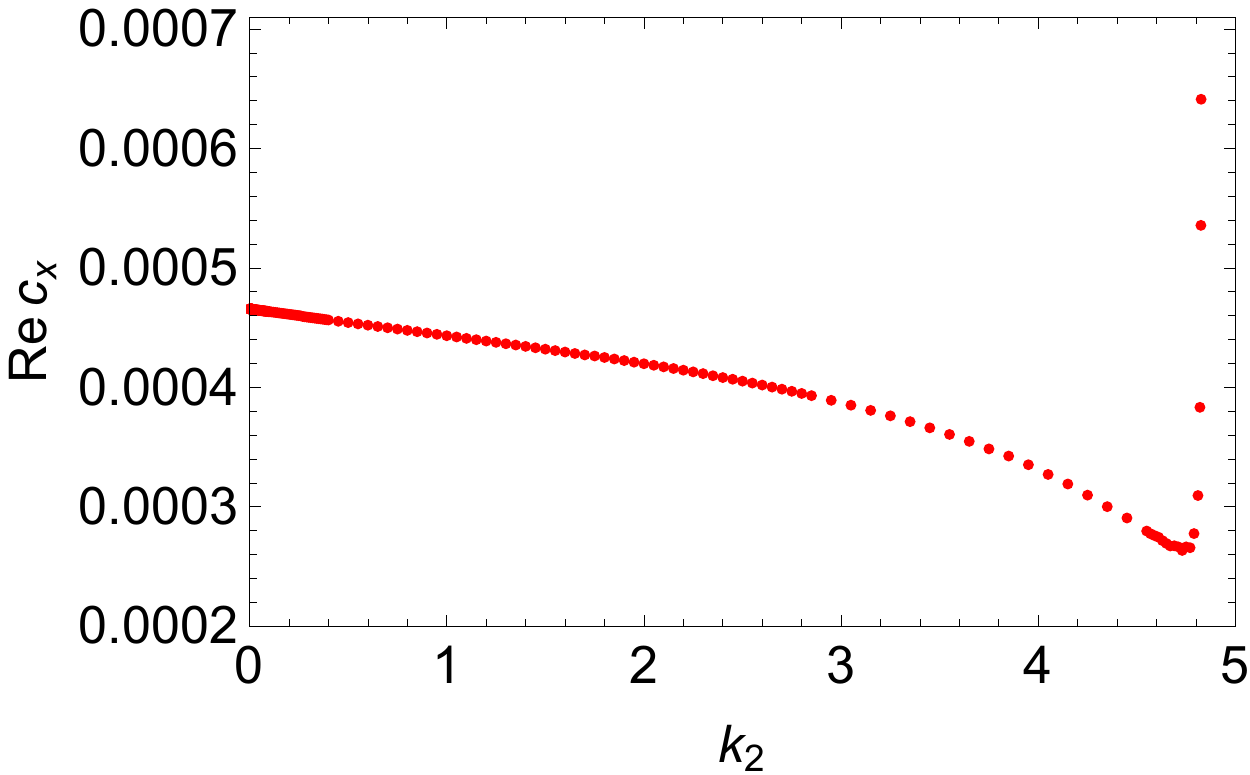}
\caption{Re$c_{x}$ in terms of $k_2$, for fixed 
$k_1=0.1$, with $\kappa=1$.}\label{Fig:cx-k2}
\end{center}
\end{figure}
\begin{figure}
\begin{center}
\includegraphics[height = 0.2\textheight]{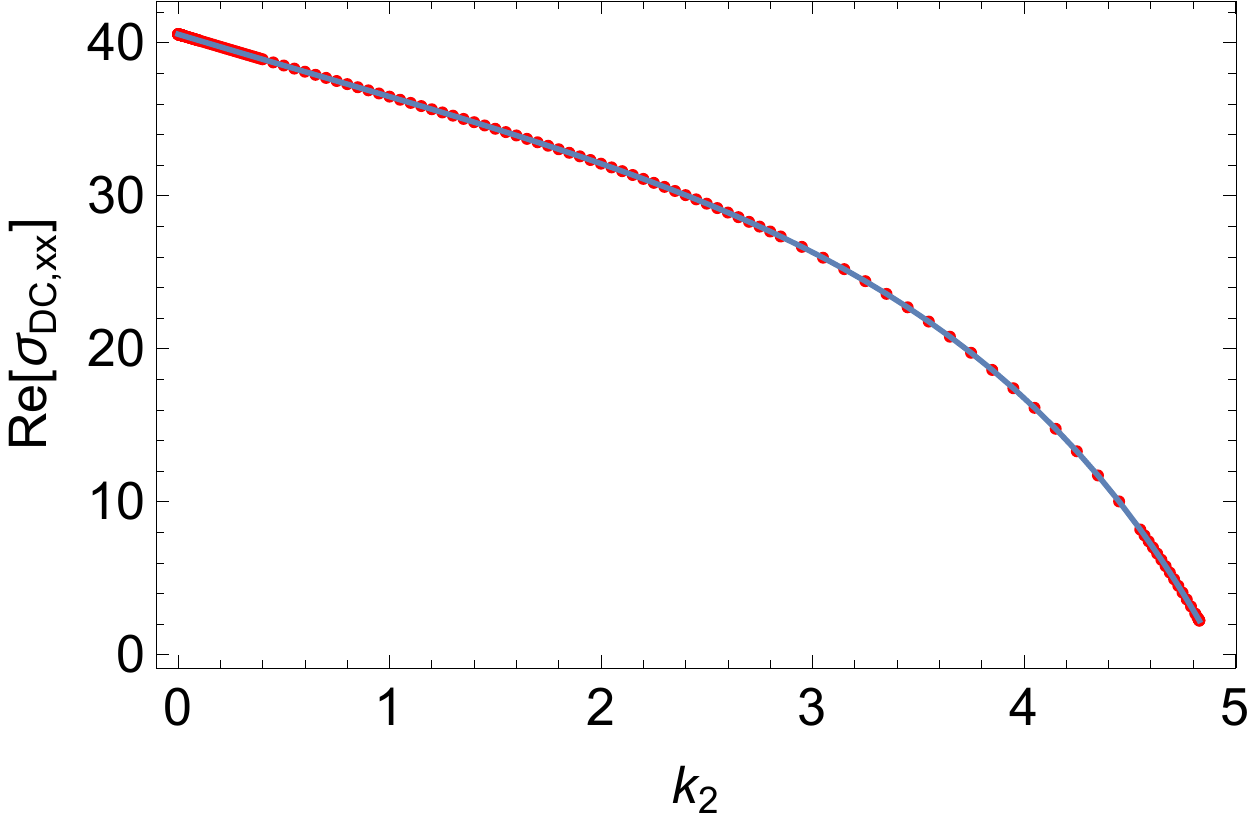}~~~
\includegraphics[height = 0.2\textheight]{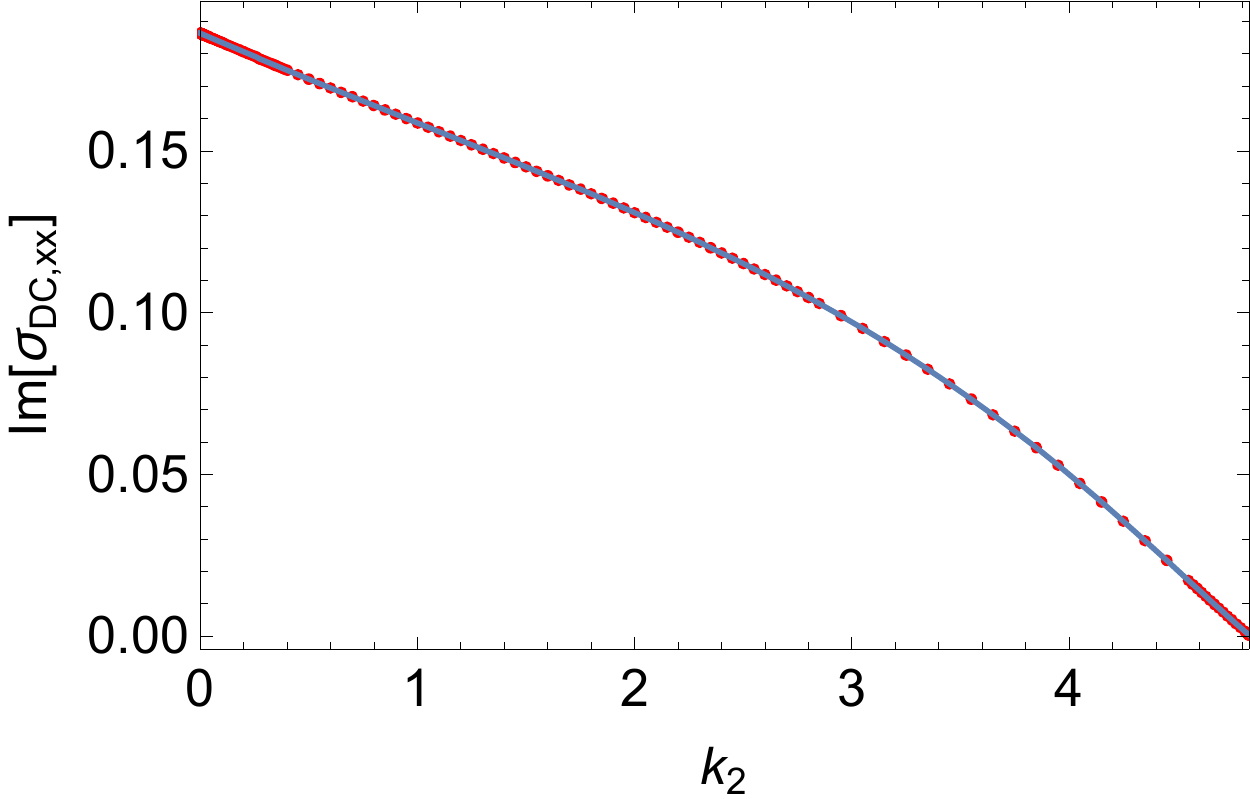}\\
~~\\
\includegraphics[height = 0.2\textheight]{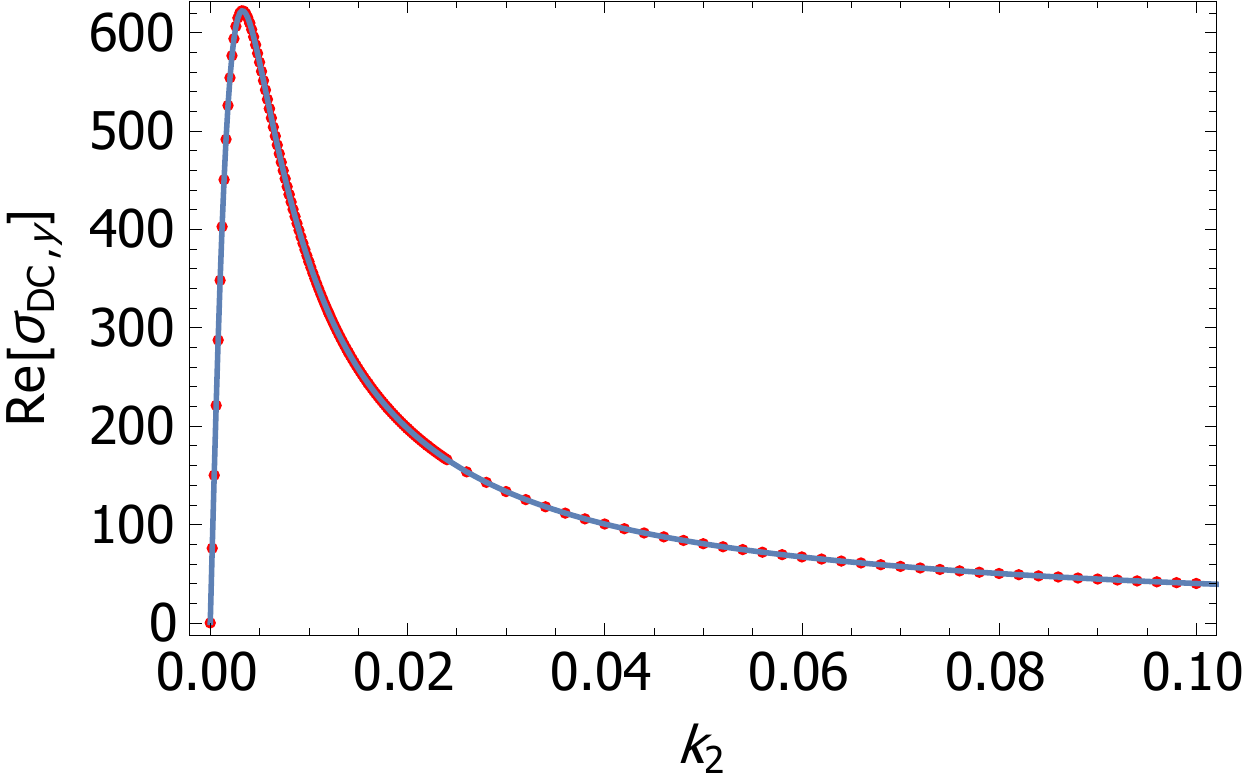}~~
\includegraphics[height = 0.2\textheight]{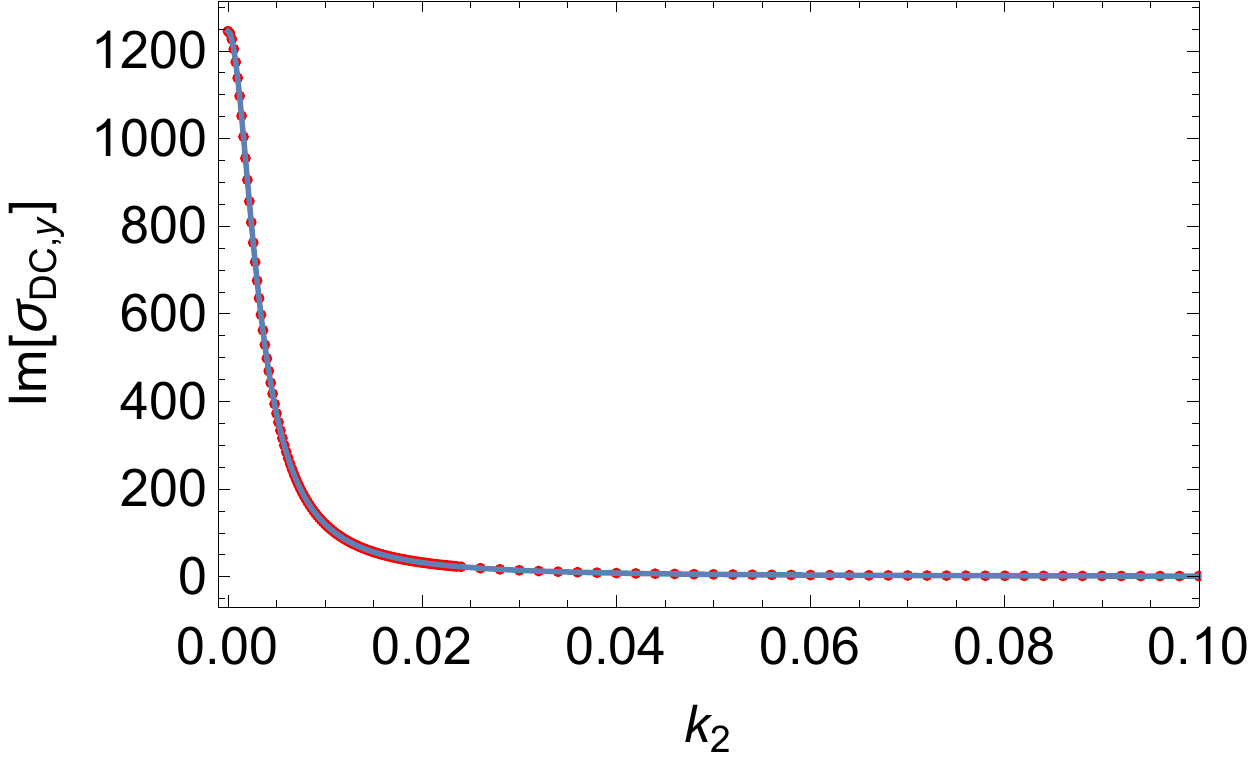}
\caption{Conductivities in terms of $k_2$,
for fixed $k_1=0.1$, with $\kappa=1$, $c_y=3.27093\times 10^{-3}$.}\label{Fig:CompAnalyticFit-k10pt1}
\end{center}
\end{figure}
\begin{figure}
\begin{center}
\includegraphics[height = 0.2\textheight]{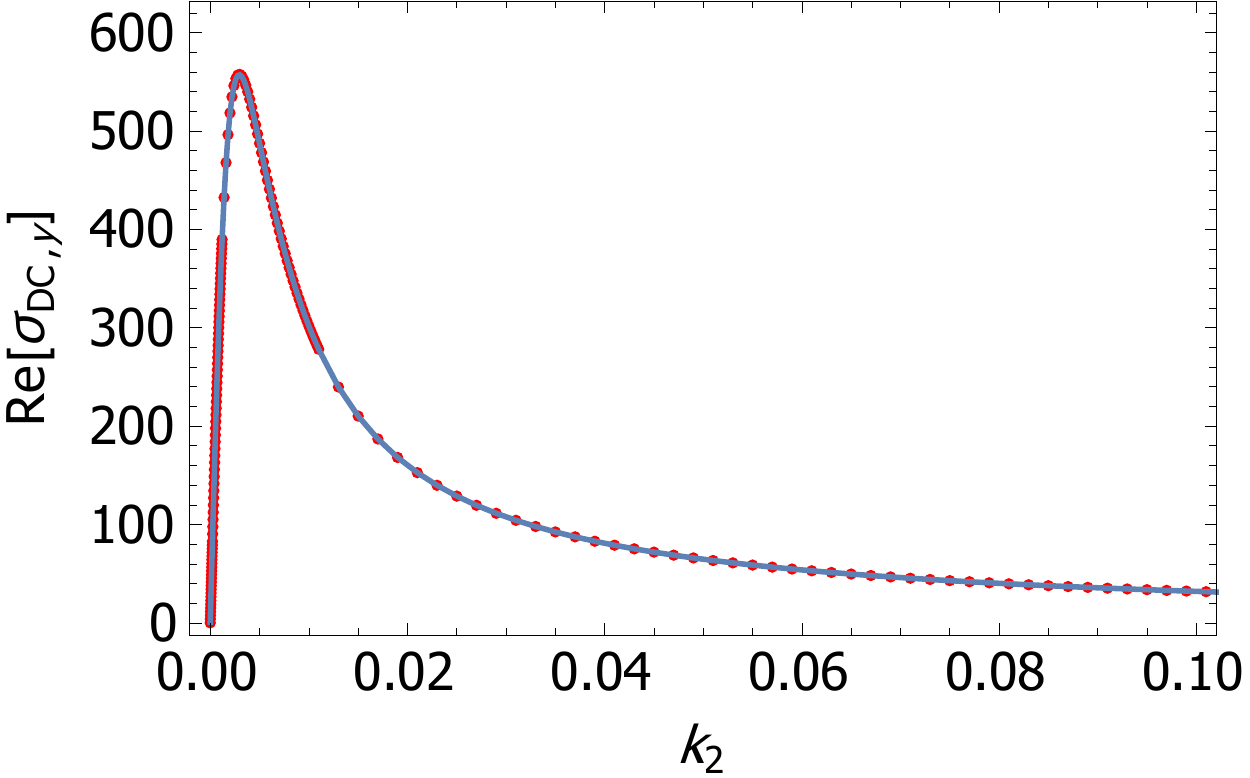}~~
\includegraphics[height = 0.2\textheight]{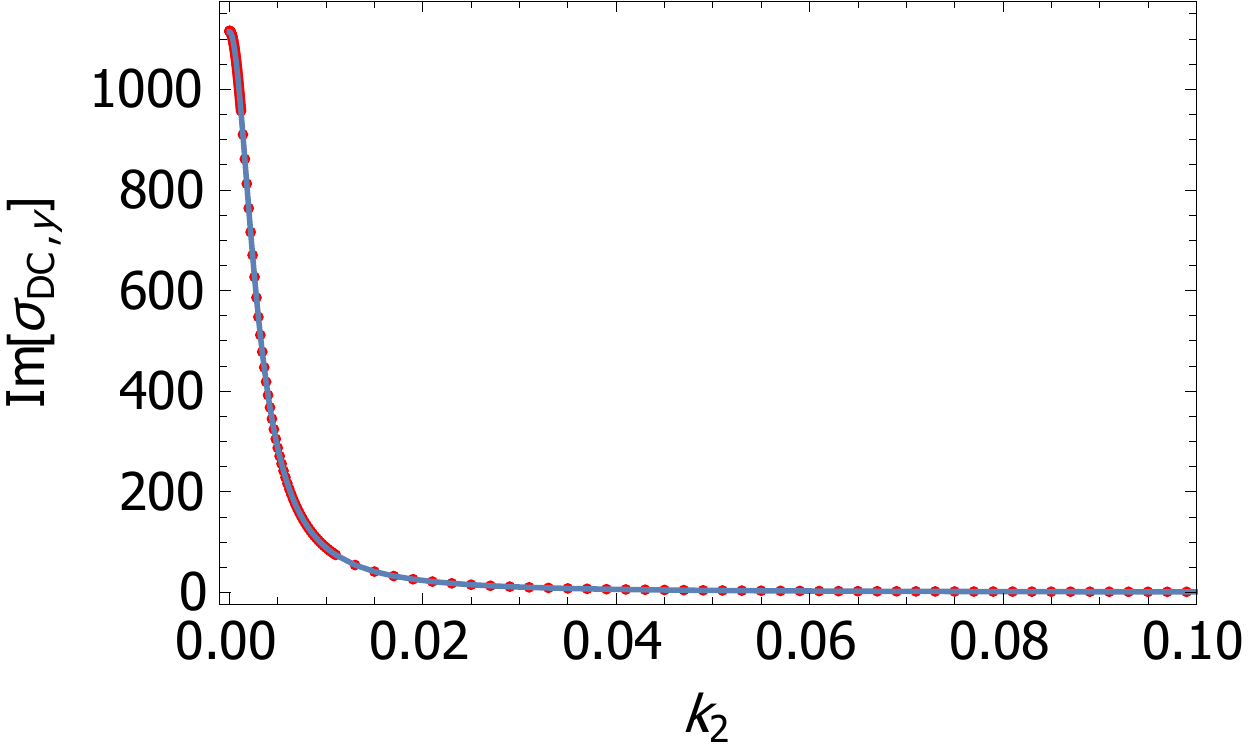}
\caption{Conductivities in terms of $\bar{\omega}$, for fixed
$k_1=2$, with $\kappa=1$, $c_y=2.9504\times 10^{-3}$.}\label{Fig:CompAnalyticFit-k12}
\end{center}
\end{figure}

{Fig.}~\ref{Fig:CompAnalytic} is the numerical results of electric conductivity by solving perturbed equations near boundary comparing with expanding solution near horizon (\ref{SigmaXXMagP20}) and (\ref{SigmaYYMagP20}) when B=0 which can be expressed as
\begin{align}
&\sigma_{Q,xx}\Big |_{z\rightarrow 1}=e^{-2 (U_B+ \phi)}-\frac{4 Q^2 e^{-(U_B+3 U_A)}}{p_1 k_1 H}\,,\label{SigmaQXX}\\
&\sigma_{Q,yy}\Big |_{z\rightarrow 1}=e^{2( U_B- \phi)}-\frac{4 Q^2 e^{U_B-3 U_A}}{p_1 k_2 H}\,.\label{SigmaQYY}
\end{align}
Above results are the exact form of DC conductivity in Einstein-Maxwell Scalar field theory done in \cite{Donos:2014cya, Andrade:2013gsa}. However, in our present anisotropic medium, we observe that (\ref{SigmaQXX}) and (\ref{SigmaQYY}) are consistent with our numerical data in large momentum relaxation strength, but inconsistent in small $k_2$ limit. However, the data plotted in {Fig.}~\ref{Fig:CompAnalytic} in $y-$direction obtained near AdS boundary is consistent with the speculating form 
\begin{align}\label{SigmaQNumYY}
\sigma_{Q,ii}\Big |_{z\rightarrow 1}=e^{2(-\epsilon_i U_B-\phi)}-\dfrac{p_1 k_i b_i H}{c_i^2+(p_1 k_i H)^2},
\end{align} 
where $b_i\equiv 4 Q^2 e^{-\epsilon_i U_B-3 U_A}$. Then, we expect that if $c_i$ is implicitly dependent on parameters $k_1$ or $k_2$, this inconsistency at $k_2\ll 1$ might be due to the effects of anisotropy. We observe that the second terms in (\ref{SigmaQXX}) and (\ref{SigmaQYY}) dominate the first terms if we input our preferred numerical data near horizon limit so that DC conductivity would be negative if $p_1>0$ (we keep $k_i\geq 0$); it is also the case for the results in AC conductivity near AdS boundary in the limit $\omega\rightarrow 0$. From now, we will present our numerical results by fixing $p_1=-1$ and $p_2=0$.  This choice of parameter values is indeed in the parameter region of stable vector mode as discussed in \cite{Alberte:2014bua}, even though our primary motivation is only to have positive DC conductivity. When $k_2$ is parametrized with $k_1$ and $\kappa$ being fixed, $c_x$ changes as given in {Fig.}~\ref{Fig:cx-k2} so that $\sigma_{DC,xx}$ perfectly match, which tells us that $c_x$ is function of $k_2$, and $c_y$ is independent from $k_2$ with constant value given in caption of {Fig.}~\ref{Fig:CompAnalyticFit-k10pt1}. However, $c_y$ changes when considered for different $k_1$ as given in the caption of {Fig.}~\ref{Fig:CompAnalyticFit-k12}, which tells us that $c_y$ is a function of $k_1$. From this, we can conclude that at fixed temperature ($\kappa=fixed$) we should expect $c_i$ in (\ref{SigmaQNumYY}) being a function of momentum relaxation, $c_i\equiv F(\delta_{ij}k_i)$. In other words, the second term in denominator in (\ref{SigmaQNumYY}) should contain both $k_1$ and $k_2$. It might be possible to manually keep parametrizing $k_1$ to see how $c_y$ behaves and extract the form of $F(k_1)$, but we do not perform it since this is not necessary for our main goal. 
In $y-$direction, when $k_2$ is large, (\ref{SigmaQNumYY}) becomes (\ref{SigmaQYY}). {Fig.}~\ref{Fig:CompAnalyticFit-k10pt1}-\ref{Fig:CompAnalyticFit-k12} show fitting of numerical data for conductivity obtained near boundary (Dotted) and horizon data from (\ref{SigmaQNumYY}).
What we have shown so far is not surprising since people have investigated this momentum dissipation by turning only one direction and DC conductivity in both directions ($x$ and $y$) should be finite, but we would like to emphasize that $c_i$ does not appear when one obtains conductivity in anisotropic medium using horizon data and resort to numerical calculation which cover whole range of momentum dissipation strength. 

Notice that (\ref{SigmaQNumYY}) is the real part of the following form
\begin{align}\label{DCGeneral}
e^{2(-\epsilon_j U_B-\phi)}+\dfrac{i b_j}{c_j-i p_1 k_j H}\,.
\end{align}
We have checked the expression (\ref{DCGeneral}) that we speculate as the general form near horizon limit when one considers anisotropic property to be well-fitted with numerical data obtained using Kubo formula near AdS boundary as long as appropriate $c_i$ is concerned.

\begin{figure}
\begin{center}
\includegraphics[height = 0.2\textheight]{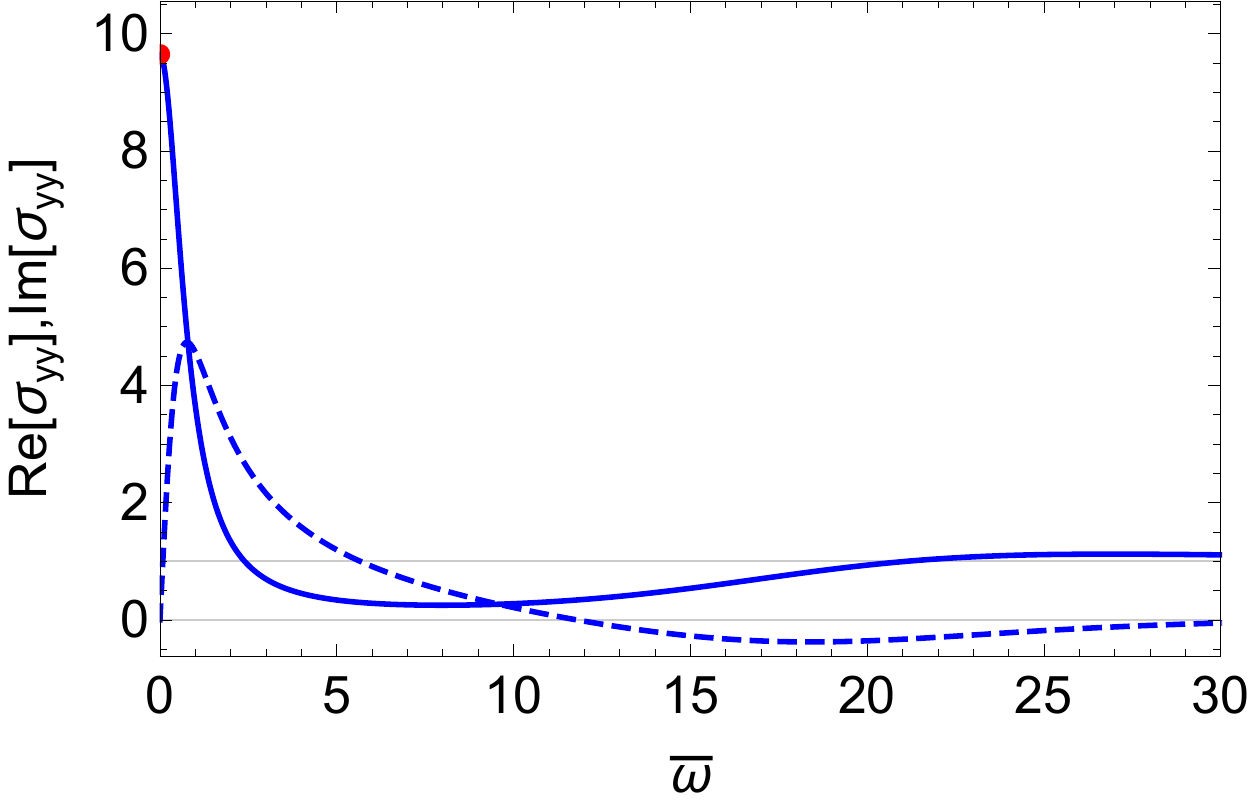}~~
\includegraphics[height = 0.2\textheight]{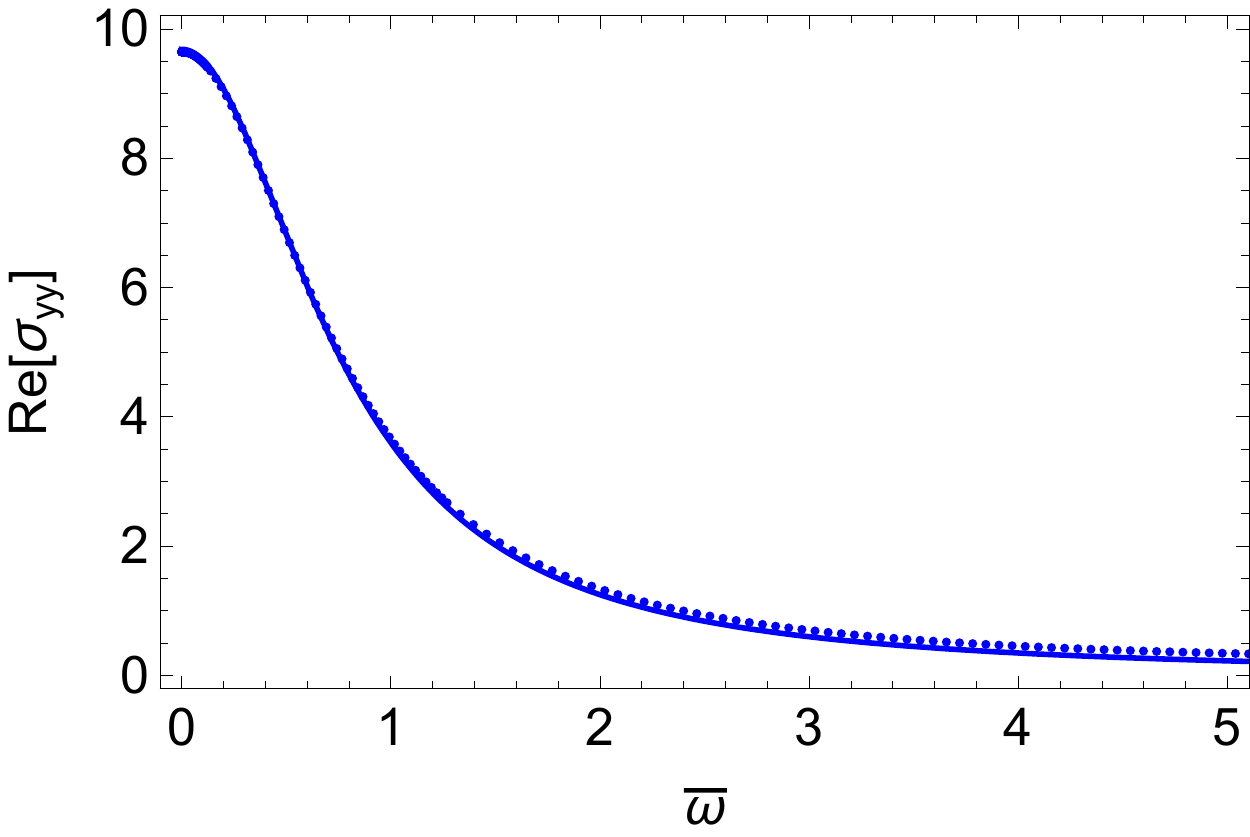}\\
~~\\~
\includegraphics[height = 0.2\textheight]{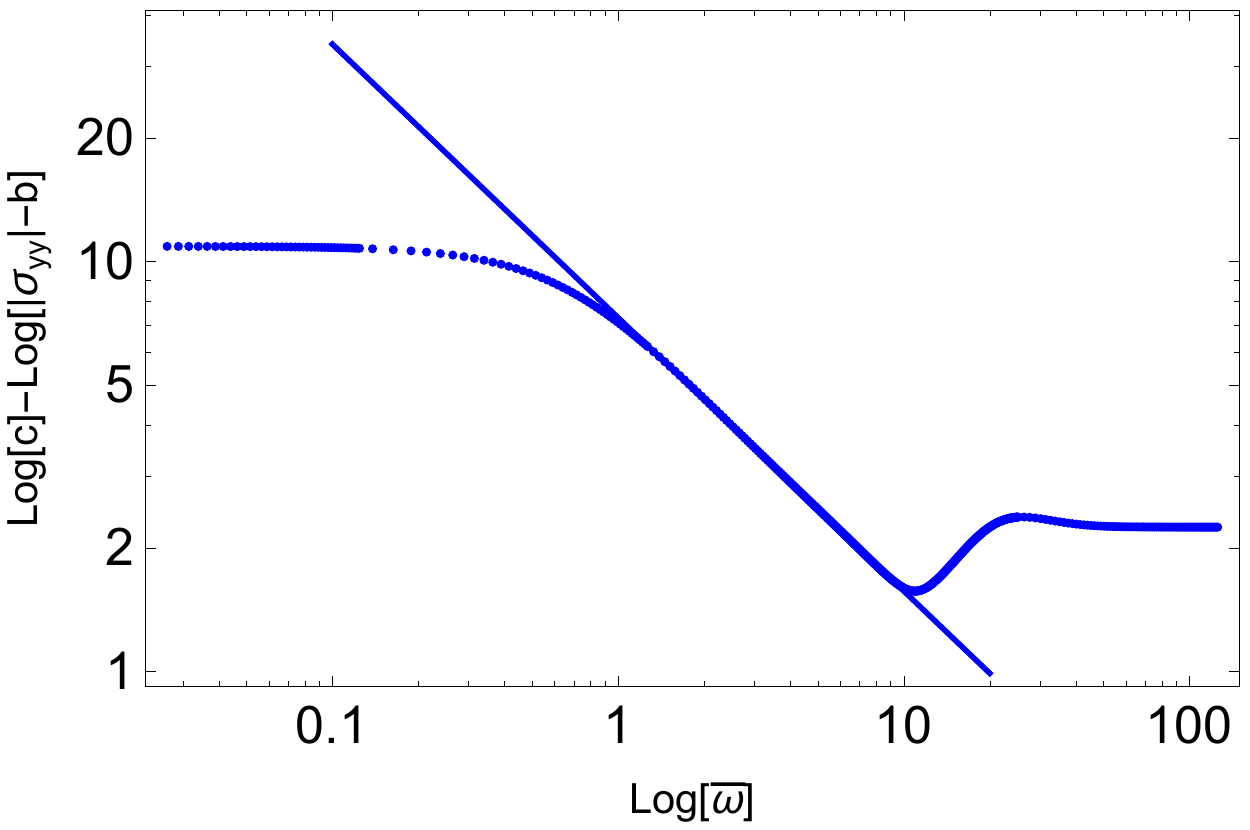}~~~~
\includegraphics[height = 0.2\textheight]{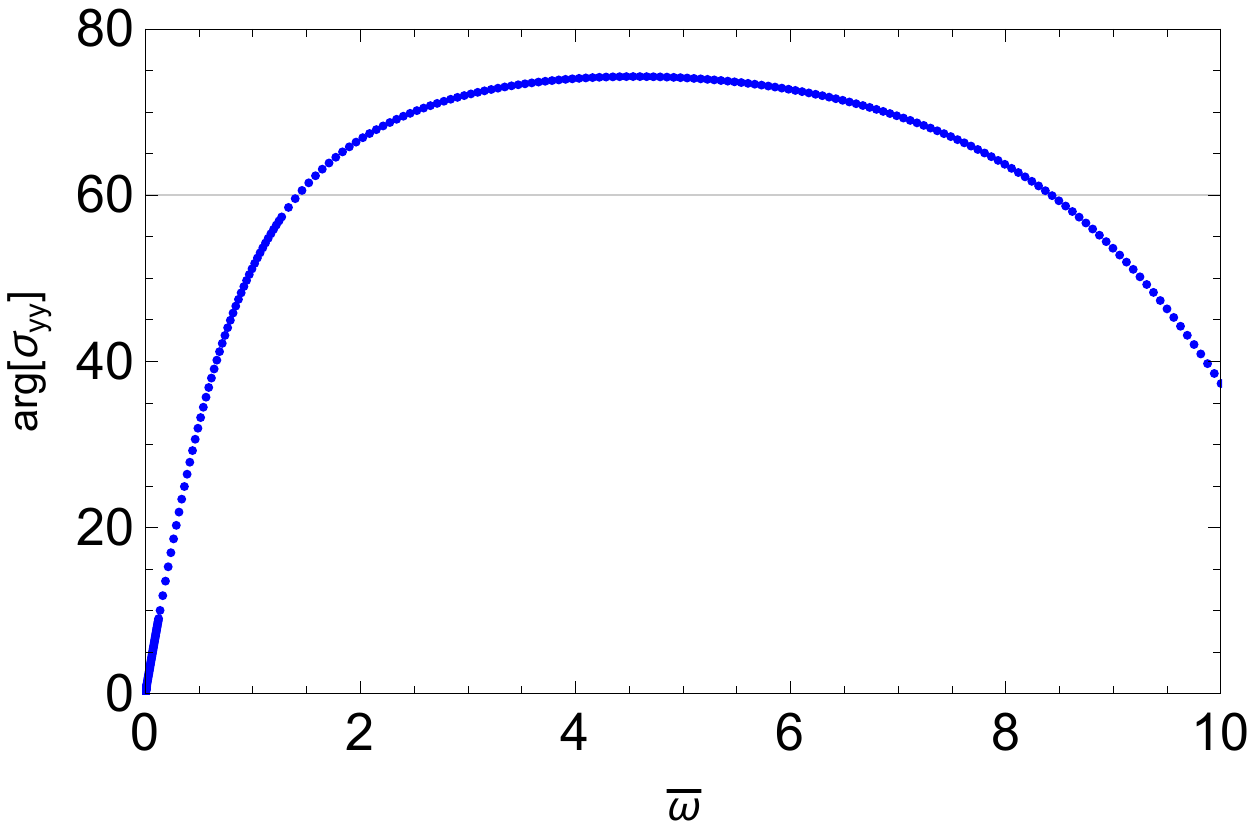}
\caption{Conductivities in terms of $\bar{\omega}$, for fixed
$p_1=-1$, $p_2=0$, $k_1=0.1$, and $k_2=0.4$.}\label{SigmaACk2-0pt4}
\end{center}
\end{figure}

\begin{figure}
\begin{center}
\includegraphics[height = 0.2\textheight]{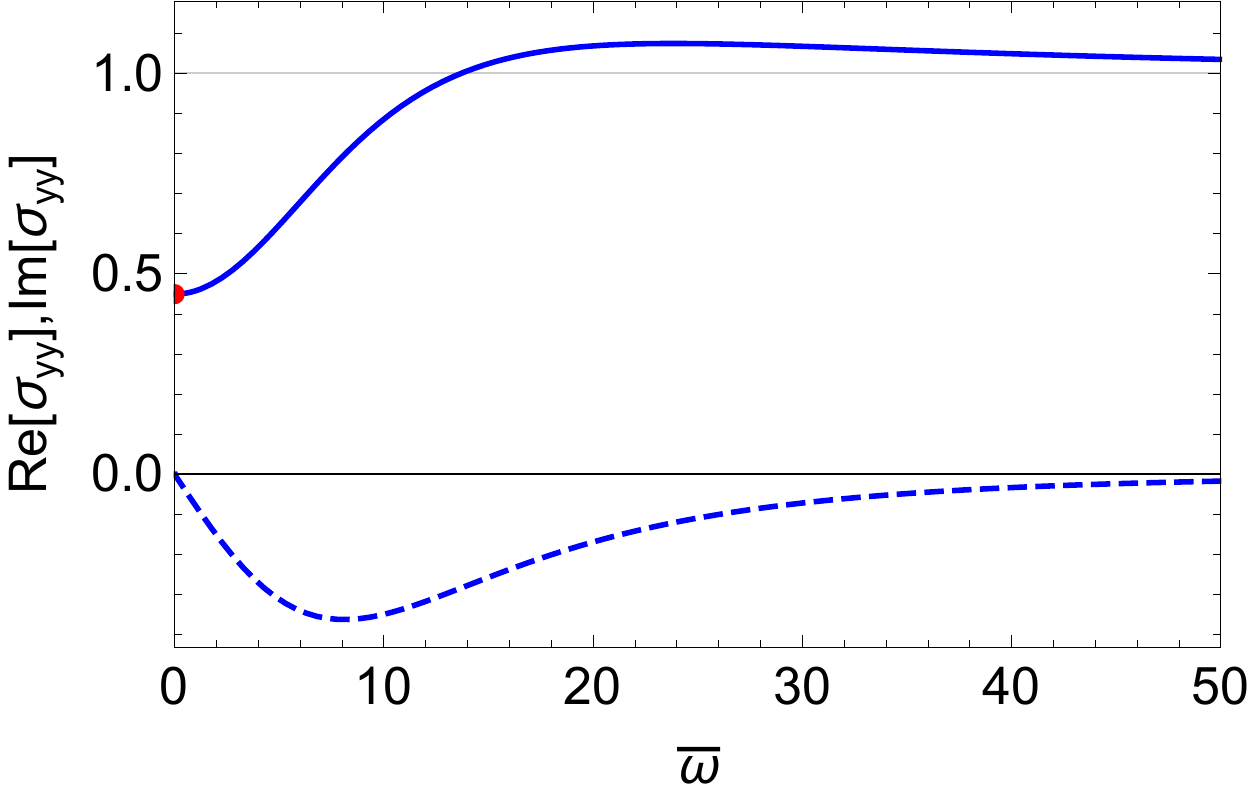}~~
\includegraphics[height = 0.2\textheight]{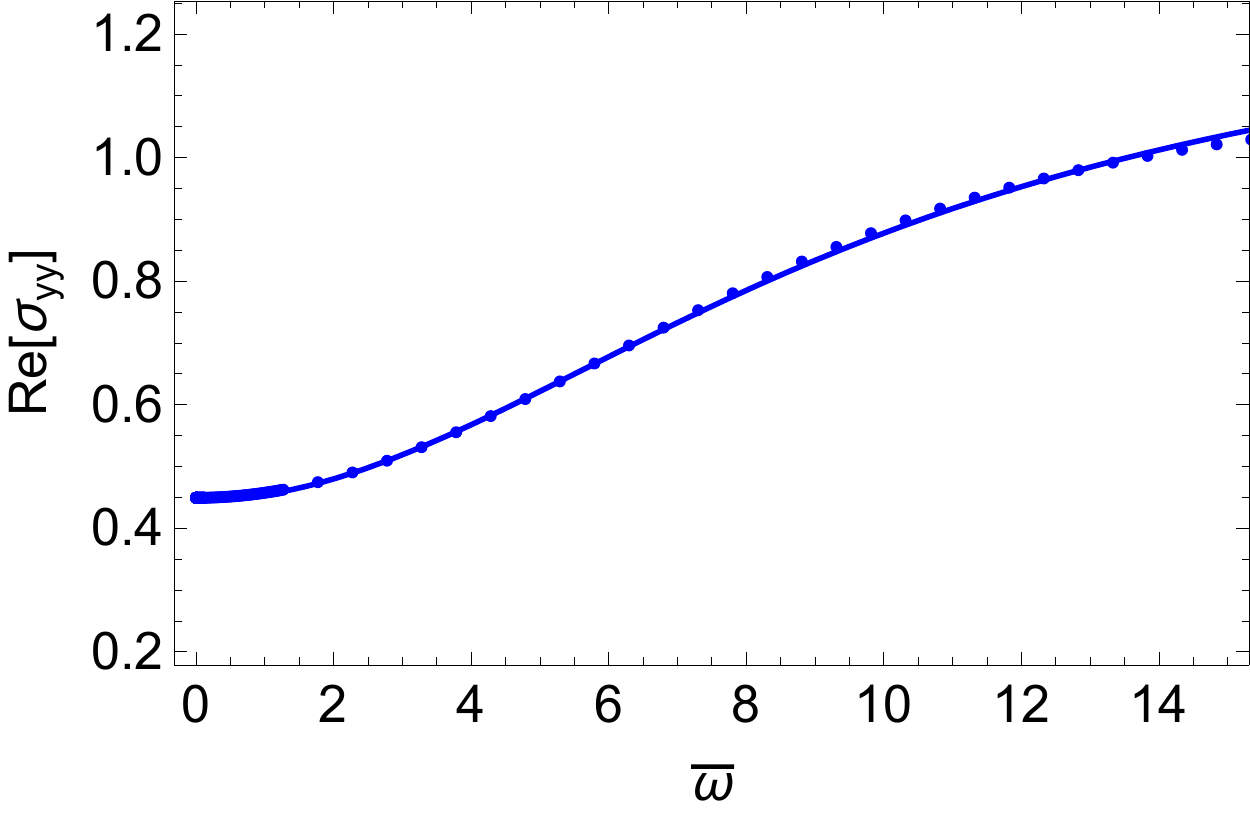}\\~~\\
\includegraphics[height = 0.2\textheight]{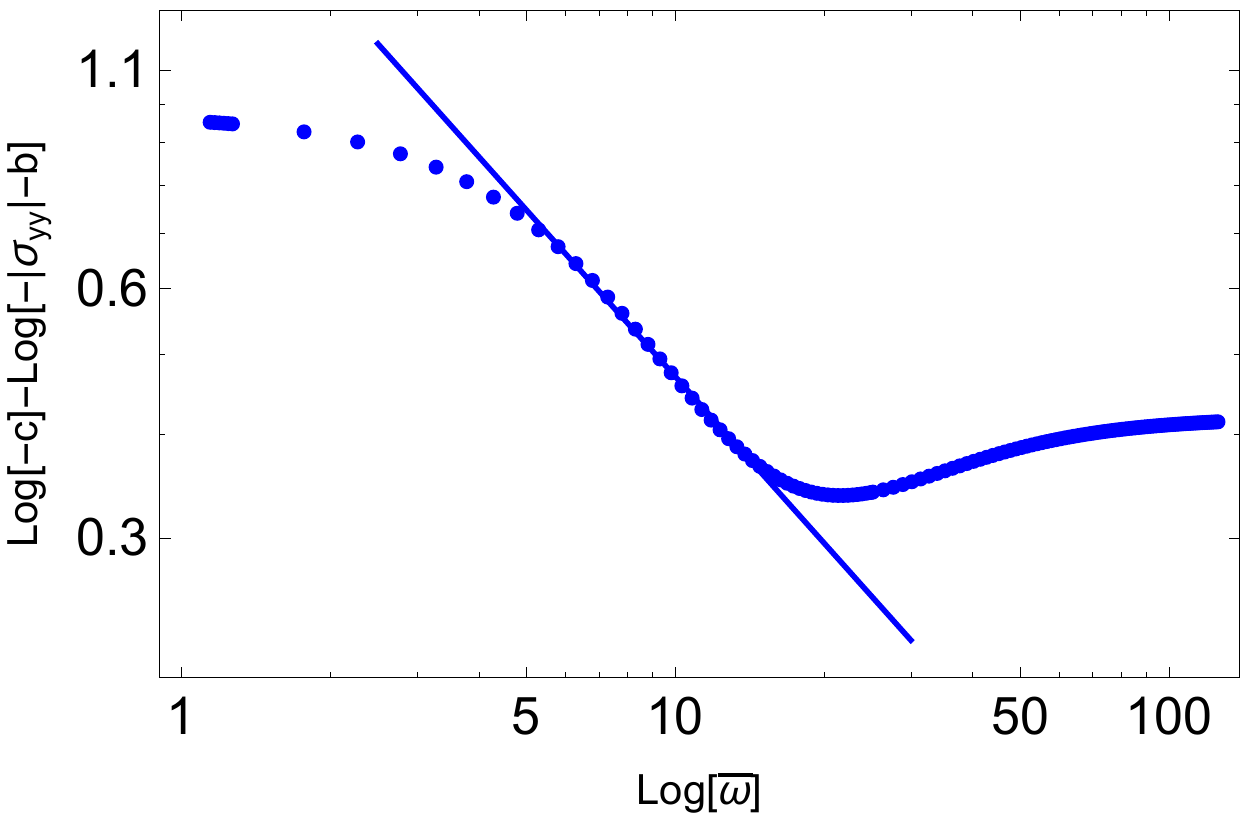}~~~~
\includegraphics[height = 0.2\textheight]{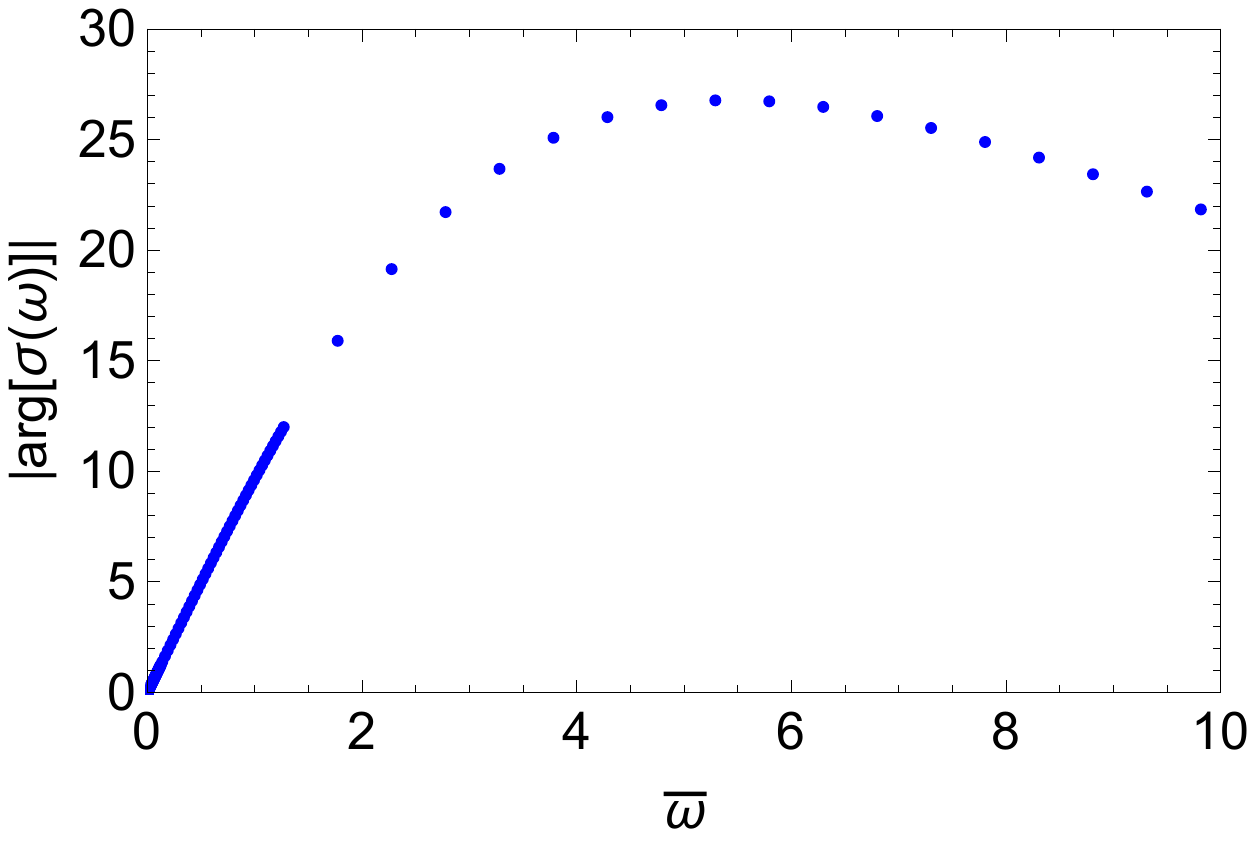}
\caption{Conductivities in terms of $\bar{\omega}$, for fixed
$p_1=-1$, $p_2=0$, $k_1=0.1$, and $k_2=4.8295$.}\label{SigmaACk2-4pt82}
\end{center}
\end{figure}

After checking AC conductivity in $\omega\rightarrow 0$ limit with results near horizon limit, we can start analyzing optical conductivity. {Fig.}~\ref{SigmaACk2-0pt4} and {Fig.}~\ref{SigmaACk2-4pt82} show the different behavior of AC coductivity due to the effect of anisotropy at low frequency for both real and imaginary part. When $k_2=0.4$, the result of real part shows the presence of Drude peak near $\omega\rightarrow 0$ while for $k_2=4.82$ Drude peak disappears.  Table \ref{Table:SigmaYY} shows the numerical values for parameters from data fitting shown in {Fig.}~\ref{SigmaACk2-0pt4} and {Fig.}~\ref{SigmaACk2-4pt82}. Drude form and intermediate scaling behavior are given as following

\begin{align}\label{DrudeForm}
\sigma(\omega)=\dfrac{k\tau}{1-i \omega \tau}+d\,,
\end{align}
and
\begin{align}\label{PowerLaws}
|\sigma_{ii}|= c\,\omega^{-\alpha}+b,,
\end{align}

\begin{figure}
\begin{center}
~\includegraphics[height = 0.2\textheight]{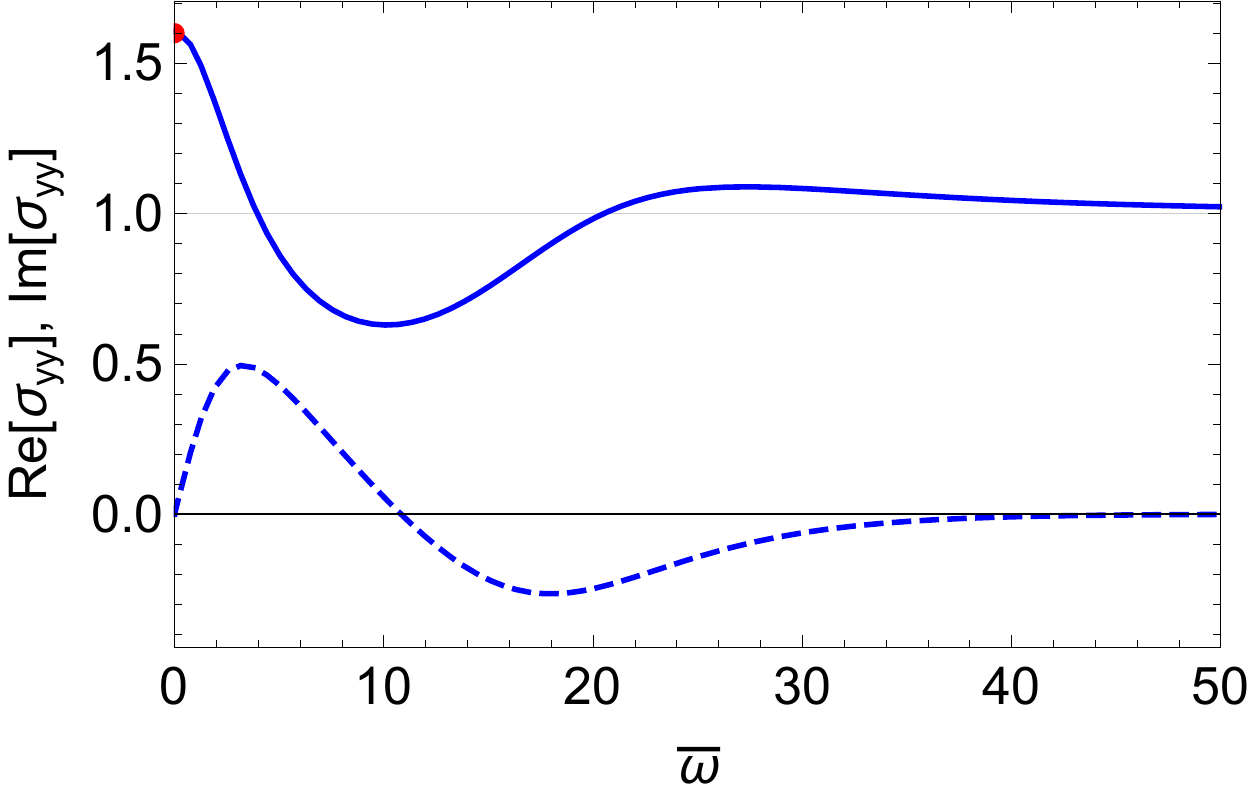}~~
\includegraphics[height = 0.2\textheight]{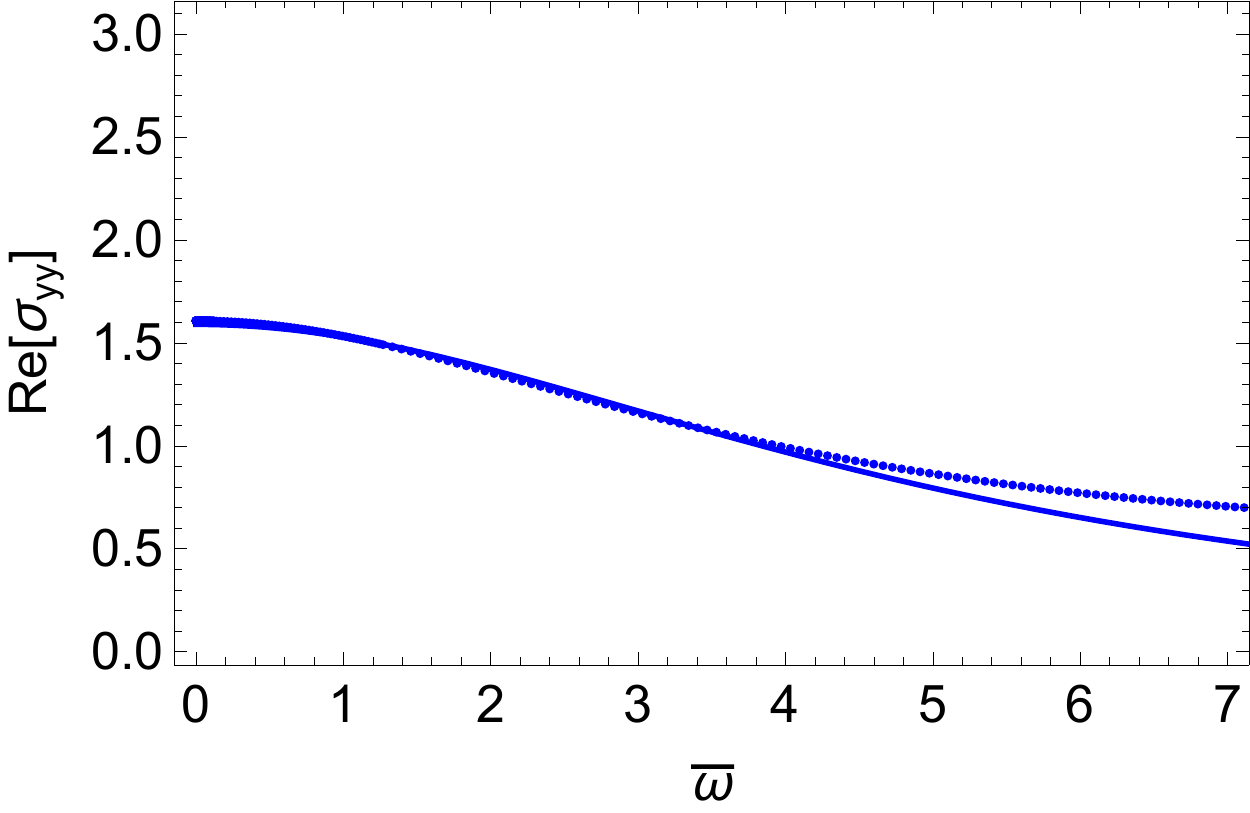}\\
~~\\
\includegraphics[height = 0.2\textheight]{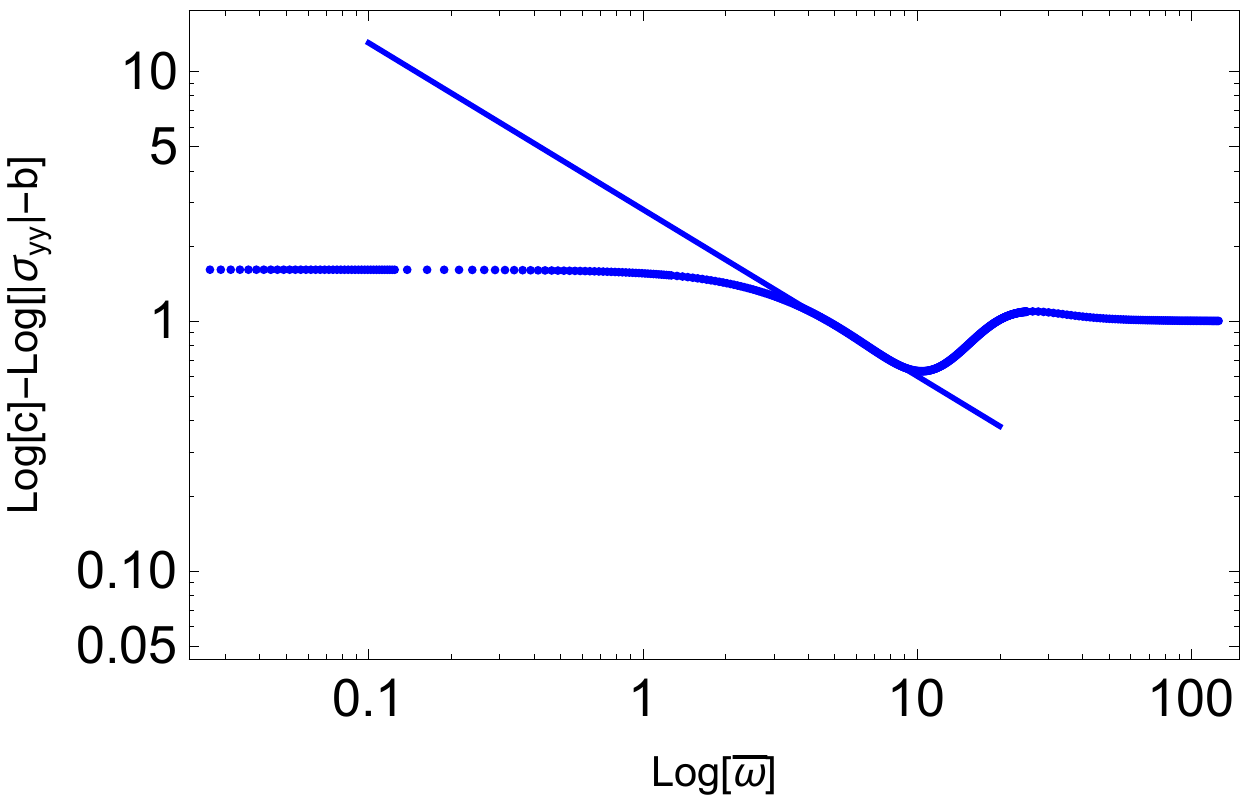}~~~~
\includegraphics[height = 0.2\textheight]{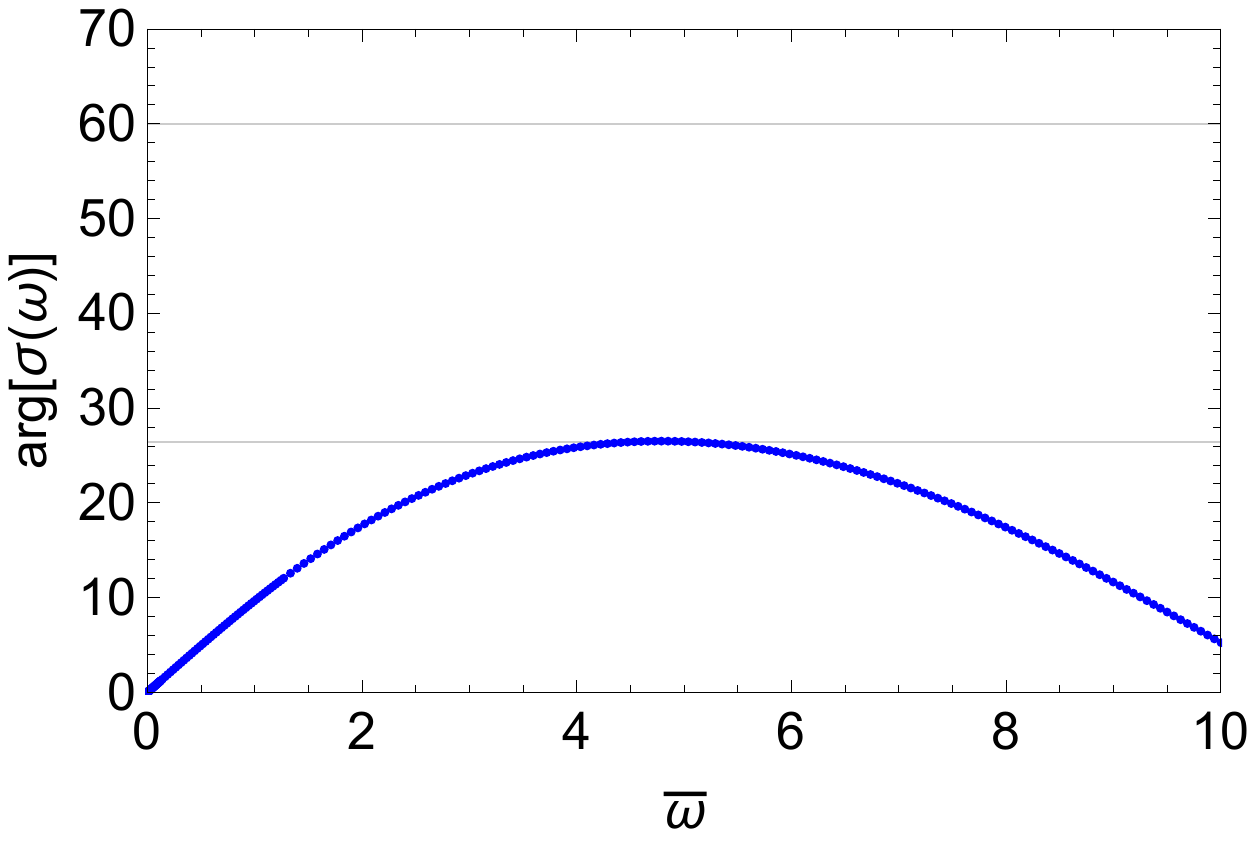}
\caption{Conductivities in terms of $\bar{\omega}$, for fixed
$p_1=-1$, $p_2=0$, $k_1=0.1$, and $k_2=2$.}\label{SigmaACk2-2}
\end{center}
\end{figure}

In (\ref{DrudeForm}) we add an offset $d$ which should be zero when AC conductivity has Drude peak. However, in some parameter values, AC conductivity in small frequency behaves as shown in {Fig.}~\ref{SigmaACk2-4pt82}, $d\neq 0$. Another interesting consequence of anisotropy in this model is that we can obtain the universal power-law behaviors for optical conductivity where $\alpha$ written in (\ref{PowerLaws}) is equal to $2/3$ for different $k_2$ which describes the property of the normal mode of cuprates, high-$T_c$ superconductor \cite{Marel:2014}. As a result, we also look for value of $k_2$ such that $b=0$ which is the case obtained by experiment. It is also quite interesting to see, at fixed temperature, how $k_1$ and $k_2$ are related such that $b=0$. From {Fig.}~\ref{SigmaACk2-0pt4}-\ref{SigmaACk2-4pt82}, we show the numerical results of $\sigma_{yy}$, the Drude peak, power-law behavior and the constant phase. We obtain conductivity using Ohm's law $\sigma(\omega)=J_j/E_j=-J_j/\dot{\bar{A}}_j=-i A_j^{(1)}/\omega  A_j^{(0)}$. Such results can be applied after putting appropriate counterterms so-called holographic renormalization \cite{ Khimphun:2016ikw, Park:2013ana, Park:2013dqa, deBoer:1999tgo, Balasubramanian:1999re, deHaro:2000vlm, Park:2014gja}. For optical conductivity, anisotropy affects the behavior of $\sigma_{yy}$ at low frequency limit.\footnote{It also affect $\sigma_{xx}$, but we do not show it here. Instead, $\sigma_{xx}$ will be used when we consider Hall conductivity.} As a result, the Drude peak for $Re[\sigma_{yy}]$ becomes flatter when $k_2$ increases and at some large values, the peak decreases its value below one as shown in {Fig.}~\ref{SigmaACk2-4pt82} where numerical factor $b,c$ and $k$ change sign. 

When $k_2=0.4$, our result in term of phase $\text{arg}[\sigma_{yy}(\bar{\omega})]$ shares similar property to massive gravity theory in \cite{Vegh:2013sk,Davison:2013jba} as well as the lattice model in \cite{Horowitz:2012ky, Horowitz:2012gs}. For $k_2=2$, we obtain offset $b=0$ so we expect to extract some analytic form of $\sigma_{yy}(\bar{\omega})$ and its phase in the range that depicts quantum critical phenomena. First, as shown in \cite{Marel:2014}, $\sigma(\omega)= c (- i \omega)^{-\alpha}=c\, \omega^{-\alpha} e^{i \alpha\pi/2}$ so that $\text{arg}[\sigma(\omega)]=90^{\circ}\times \alpha $, which is $60^{\circ}$ when $\alpha=2/3$.\footnote{Compare to \cite{Marel:2014}, $\alpha=2-\gamma$; also $\bar{\omega}=\omega/T$ } Now, our numerical result in {Fig.}~\ref{SigmaACk2-2} where $b=0$ shows that $\text{arg}[\sigma(\bar{\omega})]=26.4^{\circ}$ for $\alpha=2/3$, which suggests that electric conductivity within the region of constant phase shoud be $\sigma(\bar{\omega})= c\, \bar{\omega}^{-\alpha} e^{i (2\pi/9)\alpha}$. When we consider near critical value of $k_2$ as shown in {Fig.}~\ref{SigmaACk2-4pt82}, the peak near small frequency limit disappear with value smaller than one, and the Drude form can also be fit with additional offset $d=1.285$. We also find power-law behavior with sign-changed numerical factors except the relaxation time $\tau$ as shown in Table \ref{Table:SigmaYY}, and we then obtain $\text{arg}[\sigma_{yy}(\bar{\omega})]<0$ and plotted as absolute phase in {Fig.}~\ref{SigmaACk2-4pt82}. This is not strange to obtain negative phase due to the time reversal symmetry where $\sigma(\omega)=\sigma^*(-\omega)$ so that one should pick up the absolute phase.
\vspace{0.5cm}
\begin{table}
\begin{center}
\begin{tabular}{ | l | l | l | l |  l |  l  |  l  |}
\hline
$\sigma_{yy}$             & $\alpha$ & $c$  &   $b$ &    $k$ & $ \tau$ & $d$ \\
\hline
$k_2=0.4$                  &       2/3         & 7.3   & -1.25 & 7.45     & 1.3      & 0 \\     \hline
$k_2=2$            & 2/3               & 2.8 & 0 & 7.95       & 0.2    & 0\\      \hline
$k_2=4.8294$            & 2/3               & -2.18 & 1.42 & -8.15       & 0.103    & 1.285\\      \hline
\end{tabular} \\ 
\caption{Parameters of the Drude formula and power laws fitting the electric conductivity. }\label{Table:SigmaYY}
\end{center}
\end{table}

\begin{figure}
\begin{center}~~
\includegraphics[height = 0.2\textheight]{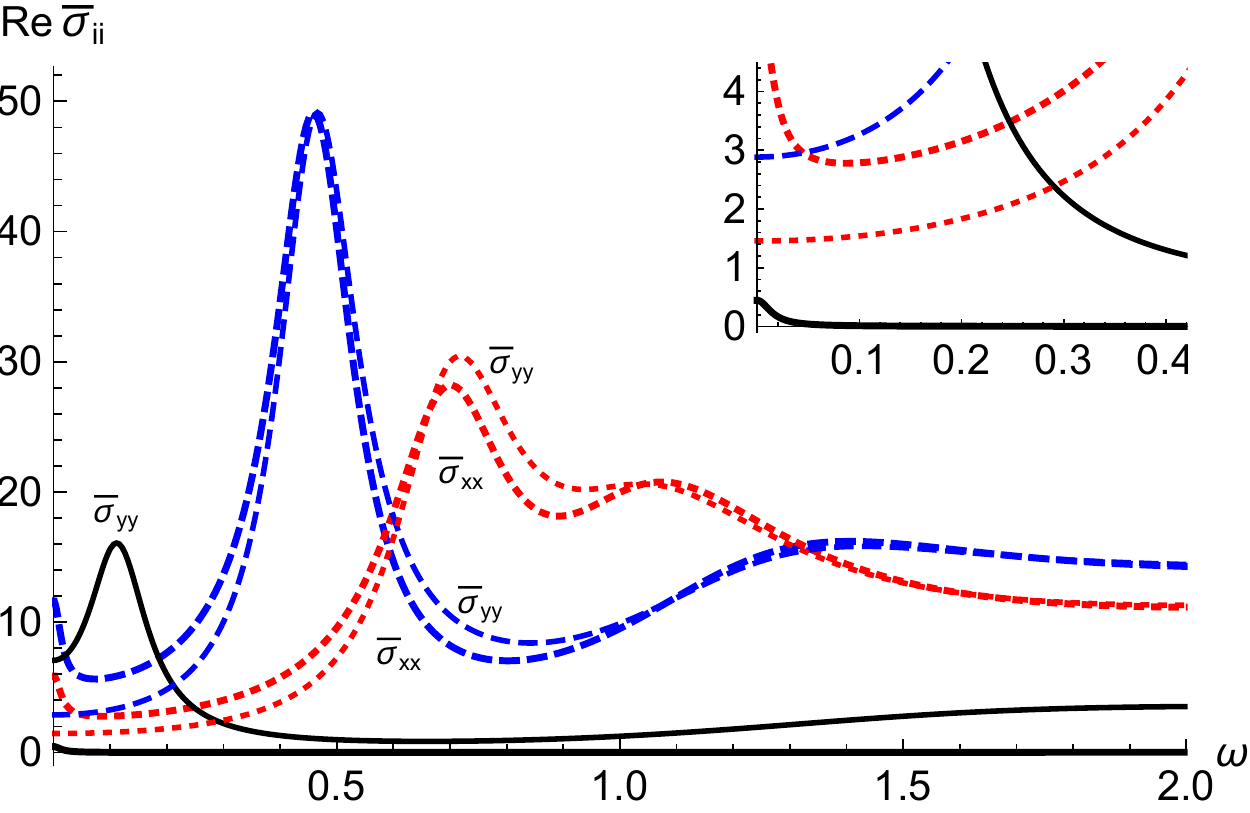}~~~~~~~~
\includegraphics[height = 0.2\textheight]{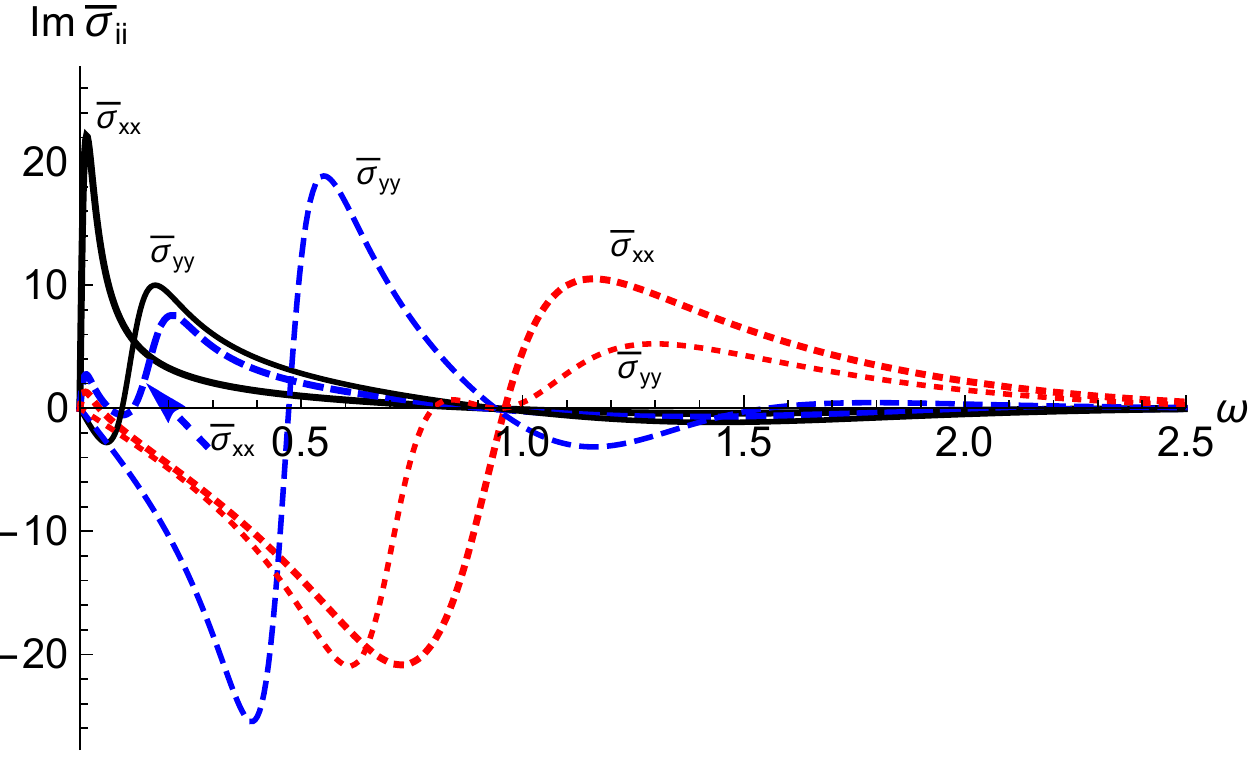}\\~~\\
\includegraphics[height = 0.2\textheight]{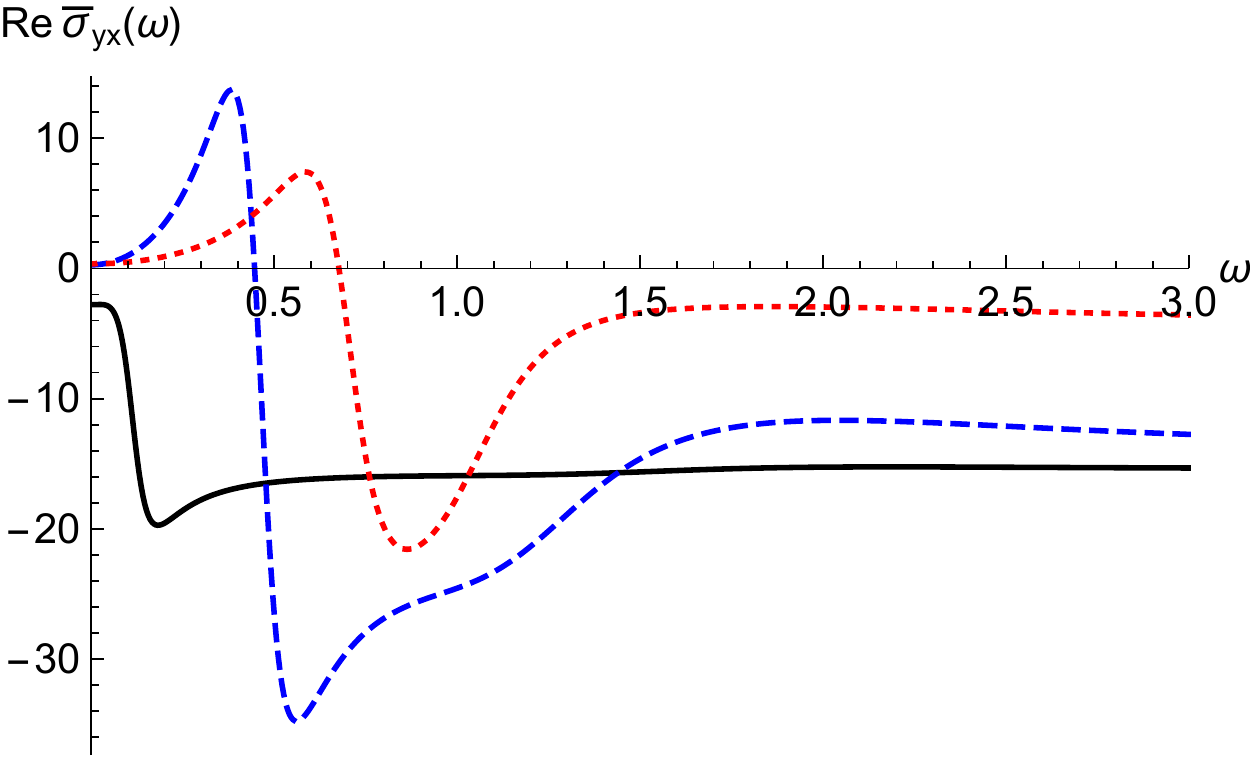}~~~~~~
\includegraphics[height = 0.2\textheight]{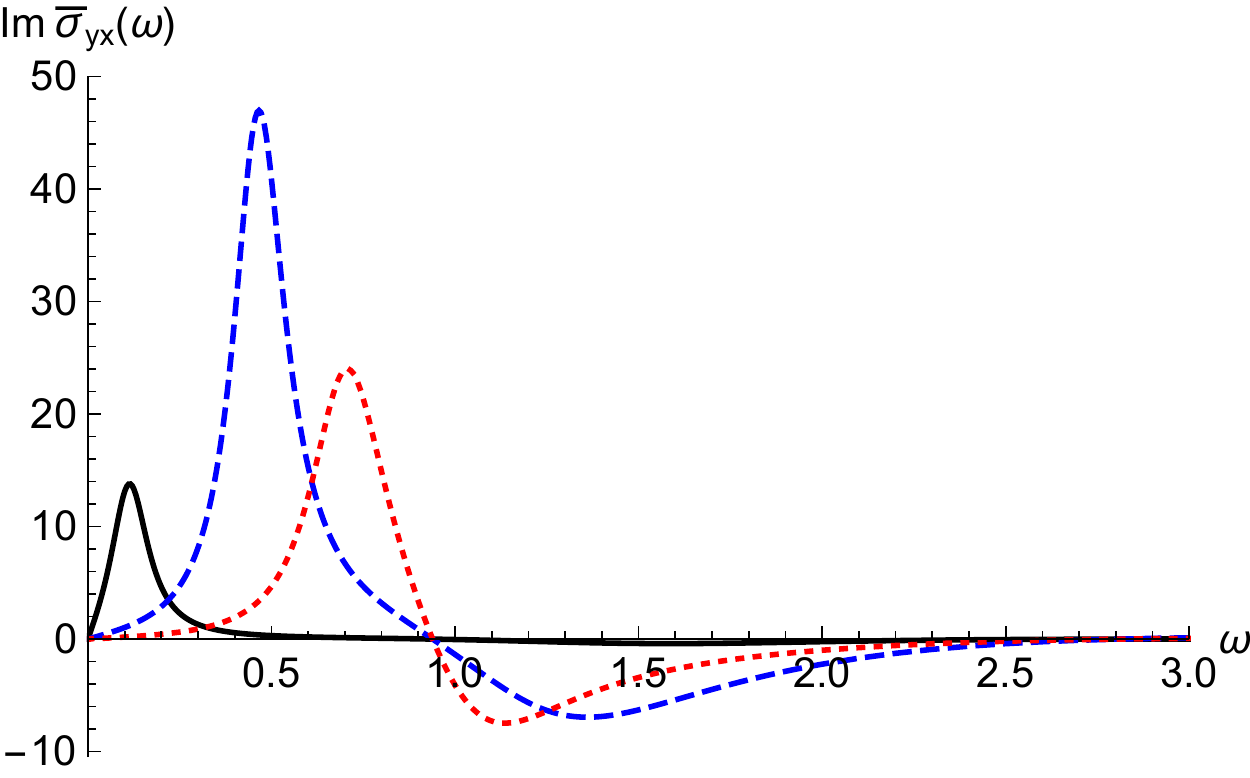}
\caption{Conductivities in terms of $\bar{\omega}$, for fixed
$k_1=0.1$ and $k_2=0.4$, $\bar{Q}_e=0.1$, with $\bar{Q}_m=0,  \sqrt{2}/2, 1$ corresponding to black-line, blue-dashed, 
and red-dotted curves.  }\label{Fig:ACSigmaBarParaQm}
\end{center}
\end{figure}

{Fig.}~\ref{Fig:ACSigmaBarParaQm} shows conductivity of EM dual by using $SL(2,R)$ transformation parametrized by $\bar{Q}_m$. As shown in {Fig.}~\ref{Fig:ACSigmaBarParaQm}, for fixed values of $\bar{Q}_e$ and $\bar{Q}_m$, we observe that $\bar{\sigma}_{xx} \neq\bar{\sigma}_{yy}$ in small frequency limit and they approach each other at large frequency. According to (\ref{SigmaBarXYGeneral}), it also implies that $\sigma_{xx}\neq \sigma_{yy}$ in small frequency limit. This behavior is due to the effect of anisotropy in $x$- and $y$-direction, but rotational symmetry remains the same as isotropic case because (\ref{SigmaBarXYGeneral}) give $\bar{\sigma}_{xy}=-\bar{\sigma}_{yx}$. The results we obtain for $\bar{\sigma}_{yx}$ in {Fig.}~\ref{Fig:ACSigmaBarParaQm} has poles and its location shifts with respect to $\bar{Q}_m$. Similar behavior should be expected when changing $k_2$ since they are both responsible for breaking translation symmetry.

\begin{figure}
\begin{center}
\includegraphics[height = 0.2\textheight]{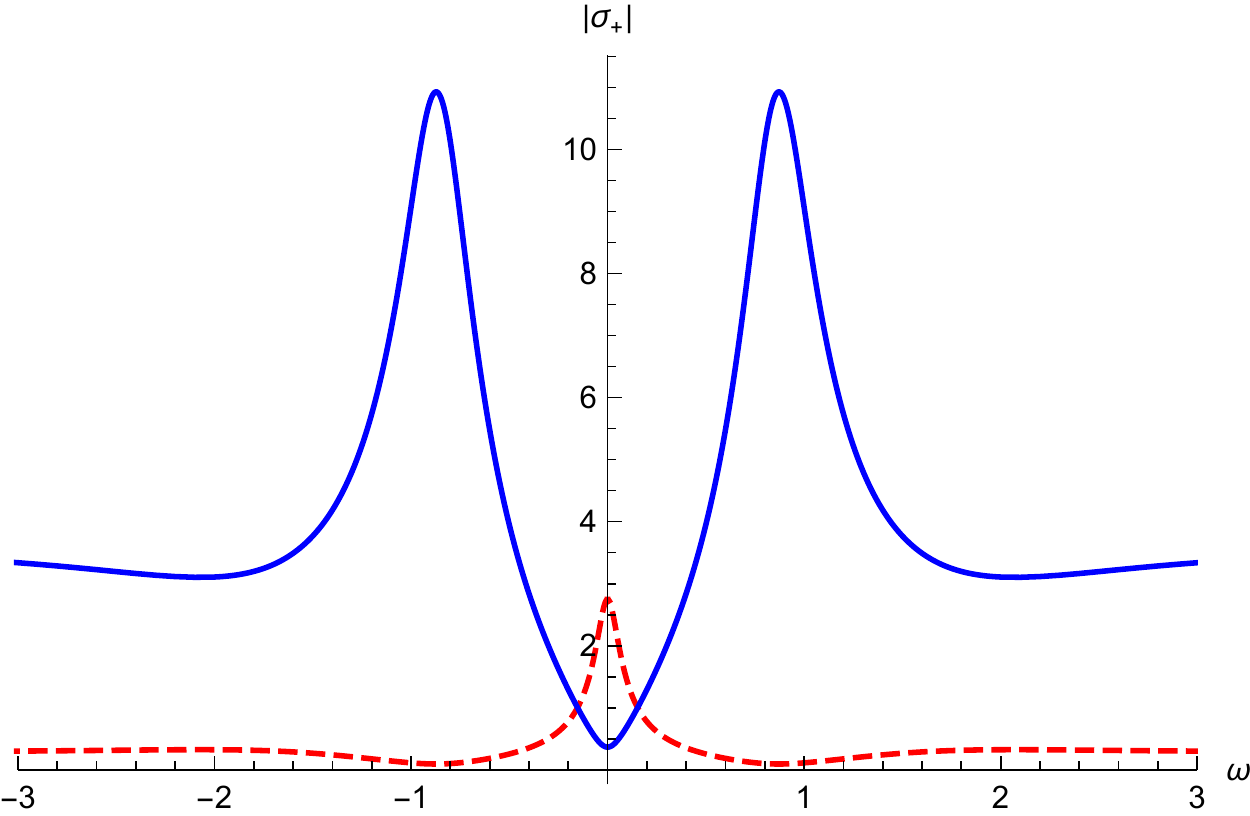}~~~~~~
\includegraphics[height = 0.2\textheight]{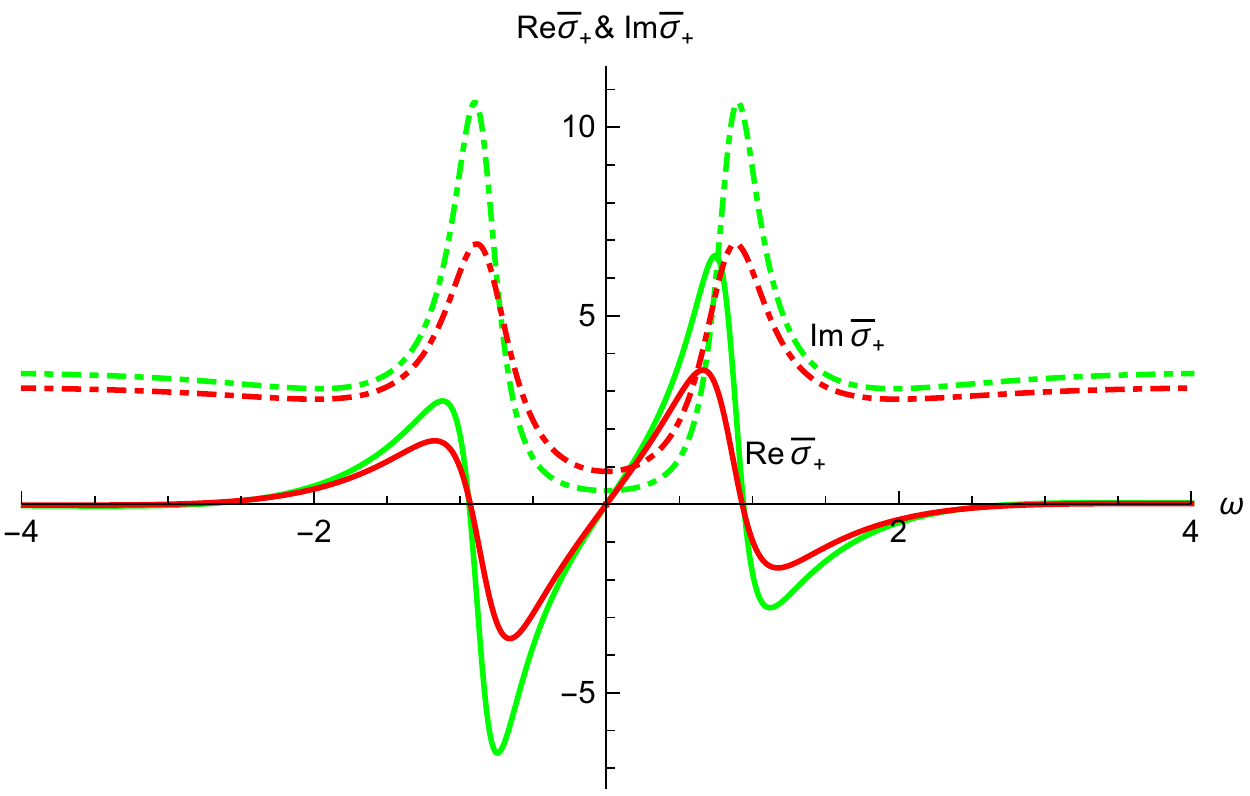}
\caption{ Left plot: Conductivities in terms of $ \omega$, with
$\bar{Q}_e=1, \bar{Q}_m=0$ (red-dashed curve), $\bar{Q}_e=0, \bar{Q}_m=1$ (Blue-solid curve) where the property of S-dual can be seen when $\bar{Q}_m\rightarrow \bar{Q}_e$ and $\bar{Q}_e\rightarrow -\bar{Q}_m$.  
Right Plot: Conductivities in terms of $\omega$, for fixed
$k_1=0.1$, $k_2=0.4$ (Green) and $k_2=1$ (Red), with $\bar{Q}_m=1$ and $\bar{Q}_m^2+\bar{Q}_e^2=1$ showing Re$\bar{\sigma}_+$ (Solid curves) and Im$\bar{\sigma}_+$ (dotted-dashed curves).}\label{Fig:SigmaBarPlusReIm}
\end{center}
\end{figure}

 There are some results in {Fig.}~\ref{Fig:SigmaBarPlusReIm} which share similar behavior with those obtained in  \cite{Hartnoll:2007ip} based on dyonic black hole is the backgound solution. First, one can check the property of S-dual for conductivity in the left plot of {Fig.}~\ref{Fig:SigmaBarPlusReIm} where we have shown that when $\bar{Q}_m\rightarrow \bar{Q}_e$ and $\bar{Q}_e\rightarrow -\bar{Q}_m$, $|\sigma_+|\rightarrow 1/|\sigma_+|$ or the blue solid curve ($\bar{Q}_e=0, \bar{Q}_m=1$) becomes red-dashed curve ($\bar{Q}_e=1, \bar{Q}_m=0$). This should be obvious since we have considered $SL(2,R)$ invariant. The right plot in {Fig.}~\ref{Fig:SigmaBarPlusReIm} are $\bar{\sigma}_+(\omega)$ defined in (\ref{SigmaPM});  for $k_2=0.4$ (green curves), the behavior of $\bar{\sigma}_+$ is qualitatively same comparing to \cite{Hartnoll:2007ip} with symmetric behavior with respect to frequency ($\sigma(\omega)=\sigma^*(-\omega)$), which is not affected by anisotropic property along $x$- and $y$-direction, but changes the phase and magnitude of this transport function.
 
\begin{figure}
\begin{center}
\includegraphics[height = 0.2\textheight]{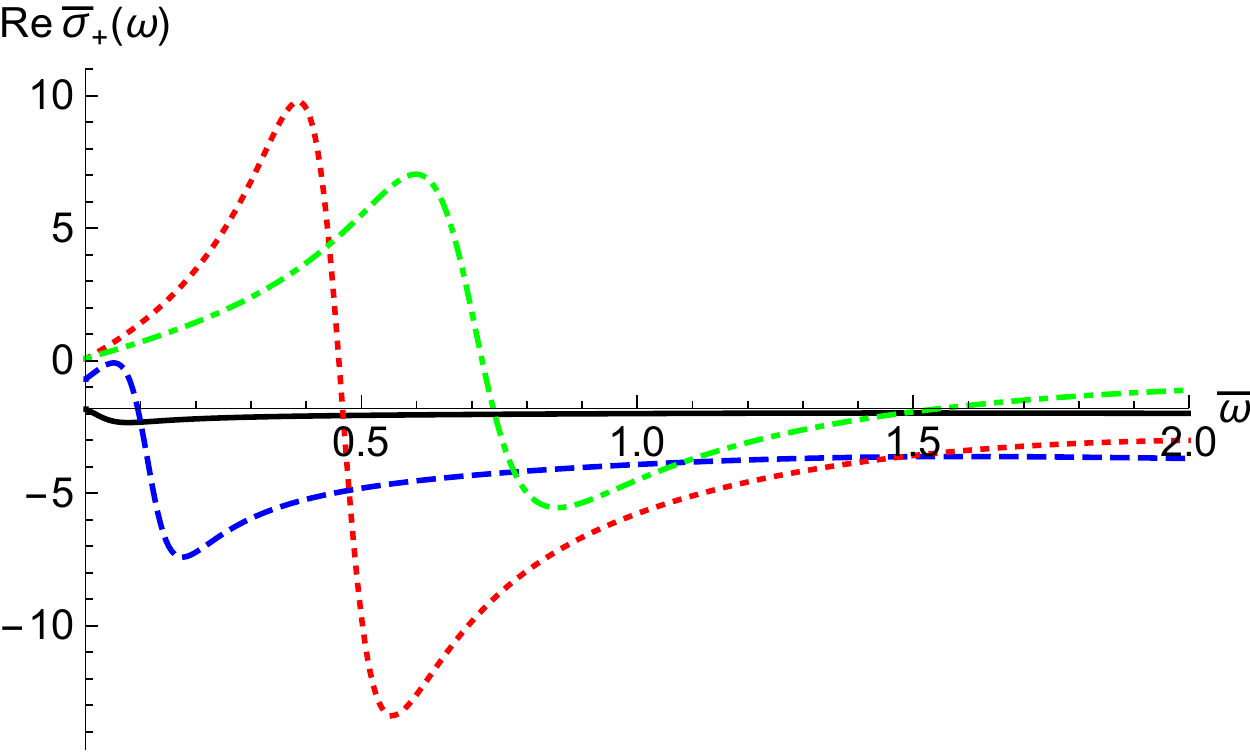}~~~~~~
\includegraphics[height = 0.2\textheight]{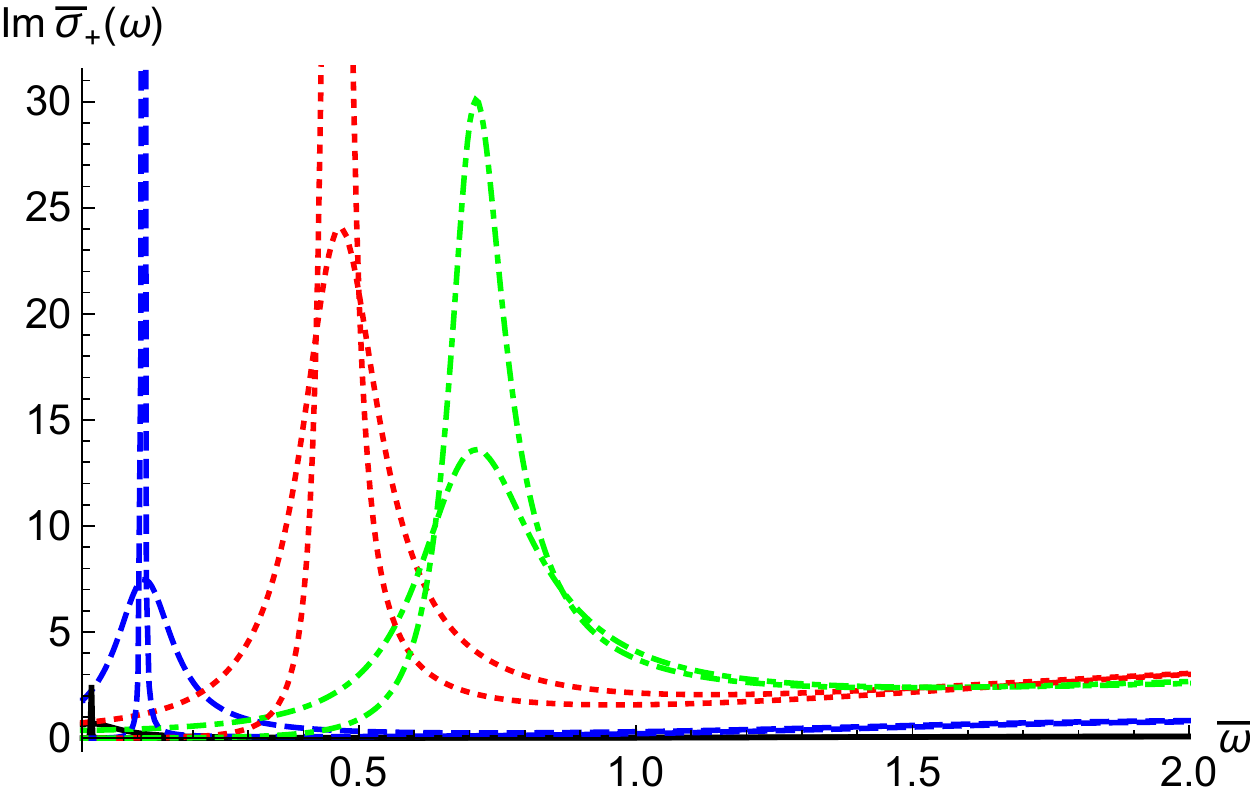}\\~~\\~~
\includegraphics[height = 0.2\textheight]{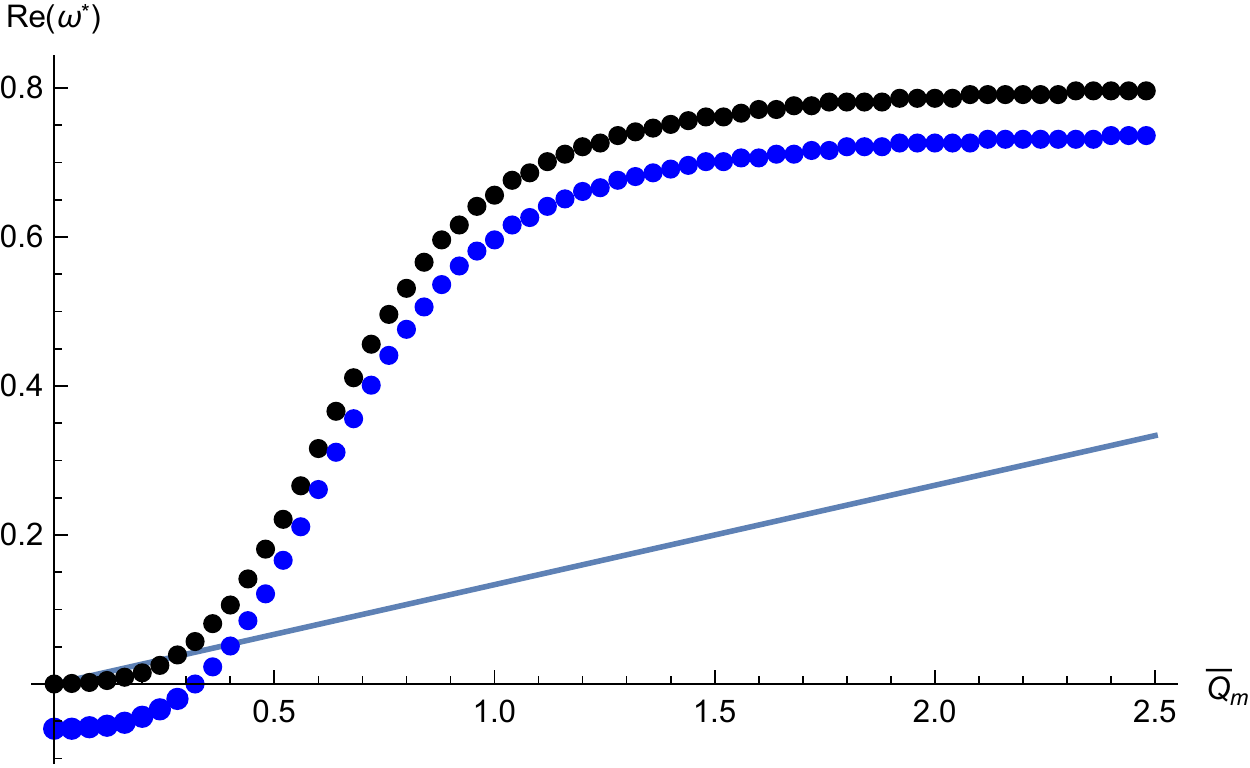}~~~~~~
\includegraphics[height = 0.2\textheight]{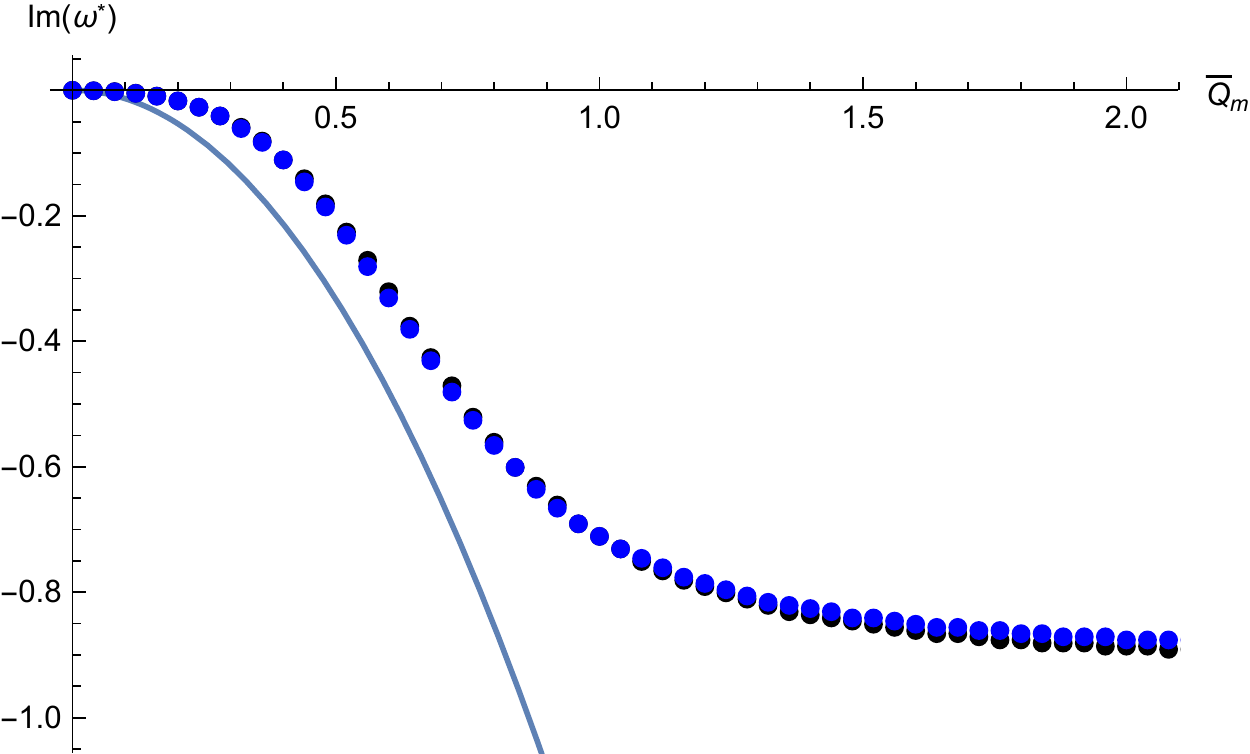}
\caption{Conductivities in terms of $\bar{Q}_m$, for fixed
$k_1=0.1$, $k_2=0.4$, and $\bar{Q}_e=0.1$. With $\bar{Q}_m=0.2, 0.4, \sqrt{2}/2, 1$ corresponding to black-solid, blue-dashed , red-dotted, and green-dotted-dashed curves, respectively. 
For dotted data in the last two plots, we show the location of the pole where $k_1=k_2=10^{-4}$ (Black) and $k_1=0.1, k_2=0.4$ (Blue). Also, the solid curves in Re$(\omega^*)$ and Im$(\omega^*)$ shows the limit of hydrodynamic of (\ref{HydroLimit}) for $\bar{Q}_e=0.1$.}\label{Fig:ACSigmaPlusParaQm}
\end{center}
\end{figure}

 We have obtained Hall conductivity in {Fig.}~\ref{Fig:ACSigmaBarParaQm} with the presence of cyclotron poles shifting with respect to $\bar{Q}_m$. As a result, in {Fig.}~\ref{Fig:ACSigmaPlusParaQm}, we plot $\sigma_+$ and pole location with respect to frequency and $\bar{Q}_m$ respectively.  This is important to compare our results with hydrodynamic limit obtained in \cite{Hartnoll:2007ip}, where $\sigma_+$ can be written in our notation as
 \begin{align*}
 \tilde{g}^2 \sigma_+= i \dfrac{4 i \tilde{Q}^2-4 \tilde{H}\tilde{Q}+3 }{4 i \tilde{H}^2+4\tilde{Q}\tilde{H}+3}\,,
 \end{align*}
 where $\tilde{Q}^2\equiv \bar{Q}_e^2/\omega $, $\tilde{H}\equiv \bar{Q}_m^2/\omega$ and $\tilde{g}^2$ is related to field theory variables as $\tilde{g}^{-2}=\sqrt{2} N^{3/2}/6\pi$. As a result, the cyclotron frequency pole can be written as\footnote{In \cite{Hartnoll:2007ip}, $\bar{Q}_e=-0.1$ which is the opposite sign with our notation due to the definition we define field strength in (\ref{Fbar}). Results in \cite{Hartnoll:2007ip} is obtained from isometric dyonic black brane and what we show in {Fig.}~\ref{Fig:ACSigmaPlusParaQm} starts from RN-AdS black brane and we obtain conductivity for EM dual from $SL(2,R)$ transformation.}
 \begin{align}\label{HydroLimit}
 \omega^*=-\dfrac{4}{3} \bar{Q}_m(-|\bar{Q}_e|+i \bar{Q}_m)\,,
 \end{align}
 where this expression is valid for small magnetic field and without momentum relaxation. Now we can return to {Fig.}~\ref{Fig:ACSigmaPlusParaQm}. The first two plots show the real and imaginary part of $\bar{\sigma}_+(\omega)$ where resonance frequency closest to the origin shifts to the right; another words, cyclotron pole increases with respect to magnetic field.

\begin{figure}
\begin{center}
\includegraphics[height = 0.38\textheight]{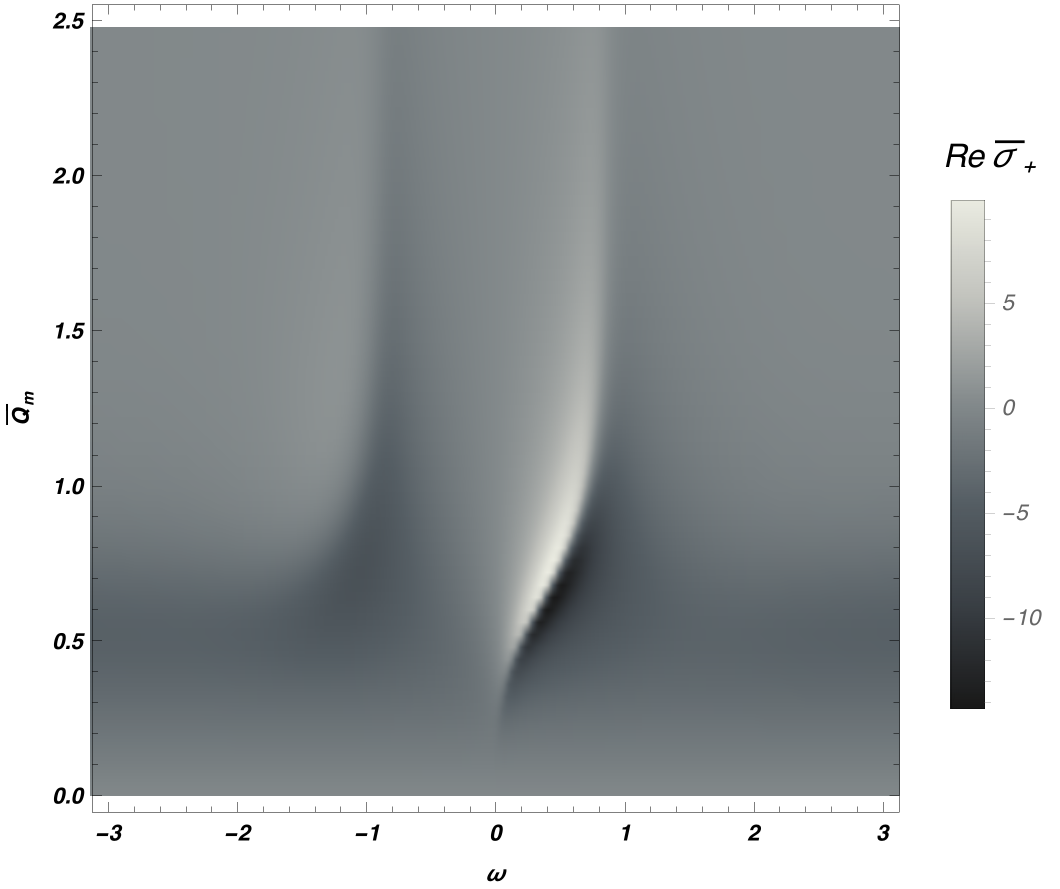}
\caption{ The density plot of Re$\bar{\sigma}_+$ in terms of $\omega$ and $\bar{Q}_m$,
 for fixed $k_1=0.1$, $k_2=0.4$, $\bar{Q}_e=0.1$. }\label{Fig:3DPlot}
\end{center}
\end{figure}

The last two plots in {Fig.}~\ref{Fig:ACSigmaPlusParaQm} illustrates the locations of cyclotron resonance with respect to $\bar{Q}_m$. In Re$(\omega^*) $ plot, blue dotted data is numerical results for $k_2=0.4$ and pole locations at small magnetic field shifts away from the straight line (Hydrodynamic limit) due to the presence of momentum dissipation and anisotropy. Then, we obtain black dotted data\footnote{
For Im$\bar{\sigma}_+$ plot, we add results for $k_1=k_2=10^{-4}$, the one with sharp peak, for double check with Im$(\omega^*)$ plot whether or not resonance frequency really shifts very slightly} for $k_1=k_2=10^{-4}$ (approaching isotropic and translation invariant), near small magnetic field limit, where poles are approaching hydrodynamic limit given in (\ref{HydroLimit}) and damped cyclotron resonance can also be observed at large magnetic field limit. In Im$(\sigma_+)$ plot, the momentum dissipation strength $k_2$ only slightly affects the pole location at large magnetic field so that near hydrodynamic limit, our data is consistent. {Fig.}~\ref{Fig:3DPlot} we show the 3D plot depicting the other possible poles  which fade out at large magnetic field.   

\newpage
\section{Conclusion}\label{Sec:Conclusion}
In this work, we have studied Einstein-Maxwell-Dilaton-Axion model minimally couple to massive gravity,
which is dual to the medium with momentum relaxation and anisotropic transports.  
Within this model we can investigate the electric conductivity and finally obtain Hall conductivity using $SL(2,R)$ invariant. 
Our primary motivation is to understand the effects of anisotropy on transport coefficients and Hall effect from EM dual using Dilatonic hairy AdS-RN backgound solutions. 
In order to consider momentum dissipation in a model with Chern-Simons term, we deviate from lattice theory, where scalar field breaks translation invariant, to non-linear massive gravity. 
We also introduce an ansatz in the reference metric (\ref{RefMetric}), which instead of constant but radial dependent, to make anisotroic solutions plausible. 

We have considered DC conductivity with background of magnetic field using horizon data and compare it with numerical results obtained near asymptotic AdS. The data matches very well in large momentum relaxation strength. Further investigation suggests that analytic solution near horizon limit should be generalized to our speculated form in (\ref{DCGeneral}), where electric conductivity $\sigma_{xx}$ and $\sigma_{yy}$ should contains both $k_1$ and $k_2$ so that both components of conductivity stay finite even though any one of the components becomes translation invariant.  

After our numerical data has passed membrane paradigm, we investigated optical conductivity and fit the data with Drude form and universal scaling behavior of cuprates. 
Electric conductivity that we obtained illustrates the dude peak behavior. The peak decreases as $k_2$ increases and at critical value of $k_2$, $\sigma_{yy}|_{\omega\rightarrow 0}<1$. In this case, to fit our data with Drude form, additional offset $d$ is needed (Table \ref{Table:SigmaYY}). Quantum critical phenomena also emerges in our model so-called universal power-law behavior which exhibits within intermediate frequency range. We found the power $\alpha=2/3$ independent of $k_2$. We also found particular case for $k_2=2$ which make offset $b=0$ so that we can see the constant phase as $80^\circ \times \alpha$ instead of $90^\circ \times \alpha$ as in \cite{Marel:2014}. 

Our final goal is to obtain Hall conductivity from $SL(2,R)$ transformation where we have initiated the electrically charged black brane instead of dyonic solution. We have obtained Hall conductivity with the presence of cyclotron resonance with damping behavior. Even though the EM duality we have considered is a special feature of CFT, our results, at a limit $k_1, k_2\rightarrow 0$, are quite consistent with MHD limit studied in \cite{Hartnoll:2007ih,Hartnoll:2007ip}. It would be quite interesting to see our cyclotron resonance non-trivially affected by magnetic field and charge density can be predicted in future observation. 

Our final comment is that it is quite interesting to investigate further the anisotropic structure of this model such that the offset in power-law scaling behavior kept to be zero in order to better understand the constant phase. Also, with Einstein-Maxwell-Scalar Field model minimally couple to massive gravity, one can consider a system inside background magnetic field, instead of using duality transformation, to see how cyclotron resonance behaves in anisotropic medium.

\vspace{1cm}
\newpage

{\bf Acknowledgement} \\

S. Khimphun acknowledges the Korea Ministry of Science, ICT and Future Planning for the support of the Visitors Program at the Asia Pacific Center for Theoretical Physics (APCTP). 
This work was supported by the Korea Ministry of Education, Science and Technology, Gyeongsangbuk-Do and Pohang City. S. Khimphun and
B.-H. Lee was supported by the National Research Foundation of Korea (NRF) grant funded by the Korea government MSIP No.2014R1A2A1A01002306(ERND). C. Park was also supported by Basic Science Research Program through the National Research Foundation of Korea funded by the Ministry of Education (NRF-2016R1D1A1B03932371).

\end{document}